\documentclass[fleqn,usenatbib]{mnras}

\usepackage[T1]{fontenc}
\usepackage{ae,aecompl}

\usepackage{graphicx}   
\usepackage{amsmath}    
\usepackage{amssymb}    

\usepackage[english]{babel}

\usepackage[multi-part-units=single]{siunitx}
\usepackage{multirow,tabularx}
\usepackage{todonotes}

\newcommand{\lcg}{low $c_{gal}$ }
\newcommand{\hcg}{high $c_{gal}$ }
\newcommand{\dsm}{\ensuremath{\Delta\Sigma_m}}
\newcommand{\dsg}{\ensuremath{\Delta\Sigma_g}}

\DeclareSIUnit[number-unit-product = {}]\yr{yr}
\DeclareSIUnit[number-unit-product = {}]\Mpc{Mpc}
\DeclareSIUnit[number-unit-product = {}]\Mpch{\si{\per\h\mega\pc}}
\DeclareSIUnit[number-unit-product = {}]\kpch{\si{\per\h\kilo\pc}}
\DeclareSIUnit[number-unit-product = {}]\pc{pc}
\DeclareSIUnit[number-unit-product = {}]\h{\textit{h}}
\DeclareSIUnit[number-unit-product = {}]\dex{dex}
\DeclareSIUnit[number-unit-product = {}]\Msol{{M_\odot}}

\defcitealias{more_detection_2016}{M16}
\defcitealias{diemer_dependence_2014}{DK14}

\title[Assembly Bias and Splashback]{Assembly Bias and Splashback in Galaxy Clusters}

\author[P. Busch \& S. D. M. White]{
Philipp Busch,$^{1}$\thanks{E-mail: pbusch@mpa-garching.mpg.de (MPA)}
and Simon D. M. White$^{1}$
\\
$^{1}$Max-Planck-Institut f\"ur Astrophysik, Postfach 1317, D-85741 Garching, Germany
}

\date{Accepted XXX. Received YYY; in original form ZZZ}

\pubyear{2017}

\begin{document}
\label{firstpage}
\pagerange{\pageref{firstpage}--\pageref{lastpage}}
\maketitle

\begin{abstract}
We use publicly available data for the Millennium Simulation to explore the implications of the recent detection of assembly bias and splashback signatures in a large sample of galaxy clusters. These were identified in the SDSS/DR8 photometric data by the redMaPPer algorithm and split into high- and low-concentration subsamples based on the projected positions of cluster members. We use simplified versions of these procedures to build cluster samples of similar size from the simulation data. These match the observed samples quite well and show similar assembly bias and splashback signals. Previous theoretical work has found the logarithmic slope of halo density profiles to have a well-defined minimum whose depth decreases and whose radius increases with halo concentration. Projected profiles for the observed and simulated cluster samples show trends with concentration which are opposite to these predictions. In addition, for high-concentration clusters the minimum slope occurs at significantly smaller radius than predicted. We show that these discrepancies all reflect confusion between splashback features and features imposed on the profiles by the cluster identification and concentration estimation procedures. The strong apparent assembly bias is not reflected in the three-dimensional distribution of matter around clusters. Rather it is  a consequence of the preferential contamination of low-concentration clusters by foreground or background groups. 
\end{abstract}




\section{Introduction}

It has long been known that, according to our standard paradigm for the formation of cosmic structure, the clustering of dark matter haloes depends strongly on their mass \citep{kaiser_spatial_1984,efstathiou_gravitational_1988,mo_analytic_1996}.
At fixed mass, large simulations of $\Lambda$CDM universes have shown that halo clustering depends in addition on a host of other properties such as formation time, concentration, spin, shape, substructure fraction and internal velocity dispersion structure \citep{gao_age_2005,wechsler_dependence_2006,gao_assembly_2007,li_halo_2008,dalal_halo_2008,faltenbacher_assembly_2010}. This additional dependence is generically called 'assembly bias', It is sensitive to the specific definition of the property considered, and it varies with halo mass in different ways for different properties. 
There is still no detailed theoretical understanding of its origin, and our inability to measure the structure of individual dark haloes directly has made it difficult
to identify observationally.

Until recently, attempts to detect an observational signal of assembly bias were
inconclusive \citep[e.g][]{yang_observational_2006,tinker_correlated_2012,wang_detection_2013,hearin_dark_2014} and controversial  \citep[e.g.][]{lin_detecting_2016}. A strong indication of assembly bias as a function of halo concentration was identified by \cite{miyatake_evidence_2016} in their study of weak gravitational lensing 
by a large sample of clusters identified in the SDSS/DR-8 photometric data. Their result was confirmed at much higher signal-to-noise by \cite{more_detection_2016}, who cross-correlated this same cluster sample with individual SDSS galaxies. In both studies, the mean projected distance of individual cluster members from cluster centre was adopted as a measure of concentration and used to split the sample into equal high- and low-concentration subsamples. Differences at large radius in the mean projected mass and galaxy number density profiles of these two subsamples then provided the evidence for surprisingly strong assembly bias, $b_{lo}/b_{hi}\sim 1.5$.

\citeauthor{more_detection_2016} also used their stacked galaxy number density profiles to search for splashback signals produced by the outer caustics defined by material that is just reaching apocentre after its first passage through the inner cluster. The caustic radius is sharply defined for spherical infall models \citep[e.g.][]{fillmore_self-similar_1984,bertschinger_self-similar_1985,lithwick_self-similar_2011,adhikari_splashback_2014,shi_outer_2016} but 
is significantly blurred, even in self-similar models, by realistic deviations from spherical symmetry \citep[e.g.][]{vogelsberger_caustics_2009}. In a $\Lambda$CDM universe, these outer caustics give rise to a sudden steepening of the spherically averaged mean density profile before it flattens out at larger radii due to  contributions from neighbouring haloes. This behaviour was studied in some detail by \cite{diemer_dependence_2014} who showed it to depend on halo mass, redshift and recent halo growth. Halo growth histories are intimately connected to their concentration, so \cite{diemer_dependence_2014} also looked for a systematic dependence of splashback signal on concentration. They found that the steepest slope
attained by the mean density profile should become shallower and the radius at which it is attained should become larger as halo concentration increases. When \cite{more_detection_2016} examined the profiles of their low- and high-concentration subsamples, however, they found the opposite ordering both in the minimum slope value and in the radius where it is attained. In addition, these radii were smaller than they expected given their estimates of cluster mass, particularly for the high-concentration subsample. Assuming that cluster galaxies trace the dark matter density profile of their host halo at these outer radii this is in conflict with the simulation results. 

The cluster sample analysed by \cite{miyatake_evidence_2016} and \cite{more_detection_2016} was based on application of the redMaPPer
algorithm \citep{rykoff_redmapper_2014} to the DSS/DR8 photometric
galaxy catalogues. As its name implies, this cluster finder uses only the non-star-forming 'red' galaxies in the catalogue. Clusters are assumed to be centred on their brightest red galaxy, and every red galaxy is assigned a probability of belonging to any particular cluster which depends on its projected distance and maximal possible redshift offset (based  on the SDSS photometry) from the cluster central galaxy. This necessarily introduces a non-negligible uncertainty in the true redshift spread among cluster members. The effect of this uncertainty on cluster properties is one of the main focuses of the current paper. Another important element of redMaPPer is the introduction of an outer cluster radius that increases slowly with the number of cluster members and is used by the algorithm to define the cluster richness and to limit the projected region over which membership probabilities are non-zero. As we shall show below, this radius, in part because of its important role in the definition of cluster concentration used by \cite{miyatake_evidence_2016} and \cite{more_detection_2016}, has a significant influence on the apparent assembly bias and splashback signals identified by these authors.

This paper is organized in seven sections. Following this introduction, \autoref{sec:methodology} describes the publicly available simulation data we use, the simplified versions of the redMaPPer and concentration estimation procedures that we apply to them, and the global properties of the resulting cluster samples.
\autoref{sec:projection} begins by demonstrating that our simulated cluster samples reproduce quite well the projected mean mass and galaxy number density profiles obtained by \cite{miyatake_evidence_2016} and \cite{more_detection_2016}, including the strong apparent assembly bias signal and the surprising concentration-dependence of the apparent splashback signal. We then investigate how this apparent success is affected by the maximum offset in depth allowed for potential cluster members, our simplified representation of the effect of photometric redshift uncertainties. In \autoref{sec:three_dimensions}, we study how well the assembly bias and splashback features measured in projection correspond to their analogues inferred from the full three-dimensional mass and galaxy distributions. \autoref{sec:projection_effects} then looks in more detail at our stacked profiles to clarify the distribution in depth of the galaxies which give
rise to the differences in mean projected galaxy number profile between low- and high-concentration clusters, while \autoref{sec:rc_influence} examines how profile shapes are influenced by the radius used by redMaPPer as the effective
limit of clusters. Finally, \autoref{sec:conclusions} gives our principal conclusions.

While we were completing the analysis for this paper, \cite{zu_level_2016} published a preprint in which they repeat the lensing analysis of \cite{miyatake_evidence_2016} 
but with the cluster sample split according to a modified definition of concentration which, as they demonstrate, is significantly less sensitive to projection effects.
With this new definition, low- and high-concentration clusters show no detectable  large-scale assembly bias. \cite{zu_level_2016} conclude, as we do below, that the strong signal in the original analysis is a result of projection effects. Our own analysis (in \autoref{sec:projection_effects}) shows explicitly how this contamination of the low-concentration clusters is distributed in depth and explains why it produces an apparently constant assembly bias signal at large projected separations.
  
\section{Methodology}\label{sec:methodology}

  Our goal in this paper is to see whether the assembly bias and splashback signals detected by \cite{miyatake_evidence_2016} and \cite{more_detection_2016} are consistent with current models for galaxy formation in a $\Lambda$CDM universe. In particular, we would like to understand the origin of the strong observed dependence of bias on cluster concentration, of the unexpectedly small scale of the detected splashback signal, and of the fact that this signal varies between high and low concentration clusters in the opposite sense to that expected both in strength and in radius. For this purpose, we need a realistic simulation of the formation and evolution of the galaxy population throughout a sufficiently large volume for our analogue of redMaPPer to identify a large sample of rich galaxy clusters.

  \subsection{Data}\label{sec:data}

     \subsubsection{Dark matter distribution}

Our analysis is based on the \emph{Millennium Simulation} described in \cite{springel_simulations_2005}. This followed structure development within a periodic box of side \SI{500}{\Mpch} assuming a flat $\Lambda$CDM cosmology with parameters from the first-year WMAP results. Although these parameters are not consistent with more recent data, the offsets are relatively small and are actually helpful for this paper since they enhance the abundance of rich  clusters in the mass range of interest. The dynamical N-body simulation followed the collisionless dark matter only, representing it with $2160^{3} \sim 10^{10}$ particles of individual mass $8.6\times 10^8h^{-1}M_\odot$ and gravitational softening length  \SI{5}{\kpch}.

      Haloes and their self-bound subhaloes were identified in 64 stored outputs of this simulation using the \textsc{subfind} algorithm \citep{springel_populating_2001}, and these were linked across time to build subhalo trees which record the assembly history of every $z=0$ halo and its subhaloes. These trees are the basis for simulation (in post-processing)  of the formation and evolution of the galaxy population. Galaxies are assumed to form as gas cools, condenses and turns into stars at the centre of every dark matter halo and are carried along as halos grow by accretion and merging. Both the subhalo merger trees and the specific galaxy formation simulation used in this paper (and discussed next) are publicly available in the Millennium Database\footnote{\url{http://www.mpa-garching.mpg.de/Millennium/}} \citep{lemson_halo_2006}.

     \subsubsection{The galaxies}\label{sec:the_galaxies}

      The particular galaxy population used in this paper was created using the semianalytic model described in detail in \cite{guo_dwarf_2011}. These authors implemented their model simultaneously on the Millennium Simulation and on the 125 times higher resolution but smaller volume Millennium-II Simulation \citep{boylan-kolchin_resolving_2009}. This allowed them to tune its parameters in order to reproduce the $z=0$ galaxy population over a very wide mass range.  In this paper we will only need to consider relatively bright galaxies, well above the limit to which results for the two simulations converge. As a result we will only use data from the larger volume simulation. We will analyse the simulation data from a single snapshot at $z=0.24$. This is the mean redshift of the clusters in the SDSS sample we compare with and is the closest snapshot to its median redshift of 0.25.
      
For all galaxies, the simulated galaxy catalogue provides positions, velocities and a range of intrinsic properties, including estimated magnitudes in the SDSS photometric bands. We restrict ourselves to galaxies with $i$-band absolute magnitude, $M_{i} < -19.43 + 5\log_{10}h$, which, for our adopted value $h=0.7$, gives $M_{i} < -20.20$. The chosen magnitude limit is very close to the one corresponding to the redMaPPer luminosity limit of $0.2L_*$ at $z=0.24$, i.e. $M_{i} = -20.25$ \citep[see][]{rykoff_robust_2012}. This selection criterion leaves us with 2,239,661 galaxies and matches that adopted by \cite{more_detection_2016} for their SDSS galaxies in order to achieve volume completeness over the redshift range, $0.1\leq z \leq 0.33$.

The next step in mimicking redMaPPer procedures is to define a class of passive or `red' galaxies. For simplicity, we require the specific star formation rate (SSFR) of model galaxies to lie below \SI{1.5e-11}{\h\per\yr}. This avoids using model colour directly which would introduce a dependence on the (uncertain) modelling of  dust effects. However, the two methods produce very similar results in practice, so the choice has has no significant effect on the analysis of this paper. 897,604 galaxies qualify as red by our criterion.
  
  \subsection{Cluster Identification and Classification}
  
Given the galaxy data described above, we wish to identify clusters using a simplified version of the scheme applied to the SDSS photometric data to generate the catalogue analysed by \cite{miyatake_evidence_2016} and \cite{more_detection_2016}. We project the simulated galaxy  and mass distributions along each of the three principal axes of the Millennium simulation to obtain three `sky' images, for each of which depth information is available for the galaxies either in real space or in redshift space. In the latter case, the line-of-sight peculiar velocities of galaxies are added to their Hubble velocities to produce
redshift space distortions (RSD). These are important when considering how the use of photometric redshifts affects the assignment of galaxies to clusters (see \ref{sec:clus_algo}). The following describes our cluster identification scheme and explains how we split the clusters into equal high- and low-concentration subsamples. 

    \subsubsection{Cluster identification algorithm}\label{sec:clus_algo}

Our cluster identification algorithm, inspired by redMaPPer, finds clusters in the projected distribution of red galaxies. Every red galaxy in each of our three projections is considered as the potential centre of a cluster. The algorithm grows clusters by adding new red galaxies    
(defined as in \ref{sec:the_galaxies}) in order of increasing projected separation until the richness $\lambda$ and the cluster radius $R_c $ reach the largest values satisfying the relation given by \cite{rykoff_redmapper_2014},
    \begin{equation}
    \begin{gathered}
     R_c(\lambda)=1.0\left(\frac{\lambda}{100}\right)^{0.2}\si{\Mpch}\label{eqn:rc}
     \end{gathered}
    \end{equation}
in physical (rather than comoving) units. Initialising with $\lambda = 1$ and $R_c(1)$,  
    \begin{enumerate}
     \item we consider as possible members the $N_g$ red galaxies which lie within $R_c$ and have a (redshift space) depth offset below $\Delta z_m$ ,
     \item we calculate $\bar N$, the expected number of uncorrelated ('background') galaxies within  $R_c$ and $\Delta z_m$,
     \item we update $\lambda=N_g-\bar N$ and $R_c(\lambda)$,
     \item we check whether the current central galaxy still has a higher stellar mass than any other cluster member, otherwise we delete it as a potential central and move to the next one, 
     \item we start the next iteration at (i) if $\lambda$  has increased, otherwise we stop.
    \end{enumerate}
 
    This process usually converges quickly and only in a few cases is it unsuccessful in finding a cluster. Note that we choose to require that the central galaxy should be the one with the highest stellar mass. Only in $\sim5$ per cent of the cases is it not simultaneously the brightest in the $i$-band, and we have checked that choosing to require instead that it should the most luminous has a negligible effect on our results. In the following we will only consider clusters with $20\leq \lambda \leq 100$, again in accordance with \cite{more_detection_2016}. 
    
   We will consider three different values for the maximal redshift-space offset allowed for cluster members, $\Delta z_m = $ \SIlist{60;120;250}{\Mpch}; the largest of these is equivalent to
projecting through the full Millennium Simulation. For comparison, the $1\sigma$ uncertainty in the photometric redshift of a single SDSS red galaxy is estimated by \cite{rykoff_redmapper_2014} to be about $90\si{\Mpch}$ at the median redshift of the observed cluster sample. The total number of clusters found (summed over the three projections) is
given in \autoref{tab:cluster_p.pdf}.
    
    \begin{table}
      \caption{The size of simulated cluster samples for different maximal depth offsets, $\Delta z_m$.}
      \centering
      \label{tab:cluster_p.pdf}
      \begin{tabular}{lccr}
        \hline
        \multirow{2}{*}{Sample Name} & $\Delta z_m$ & No. Members\\
        & \si{\Mpch} & \\
        \hline
        CS60 & 60 & 9196 \\
        CS120 & 120 & 9213 \\
        CS250 & 250 & 8930 \\
        \hline
      \end{tabular}
    \end{table}
 
These numbers are similar to the number of clusters (8,648) in the
observed sample we are comparing with.  This is a coincidence since the volume of the Millennium Simulation is only about a tenth of that in the SDSS footprint over the redshift range $0.1 \leq z \leq 0.33$, but the abundance of rich clusters is enhanced by a factor of about three in the simulation because it assumes $\sigma_8 = 0.9$, significantly above current estimates\footnote{We checked the results of this paper using the public semianalytic catalogue of \cite{henriques_galaxy_2015} which is implemented on a version of the Millennium Simulation rescaled to the Planck 2013 cosmology \citep{planck_collaboration_planck_2014}. We find far fewer clusters: 2407, 2244 and 2307 for the equivalents of CS250, CS120, and CS60, respectively. This corresponds to 83.1\%,  77.5\% and  79.6\% of the expected number of clusters in three times the (rescaled) volume of the simulation. We decided to stay with the original cosmology since the larger number of clusters provides much better statistics.}.

There is, of course, a very substantial overlap between these three cluster samples, but it is not perfect. In \autoref{tab:cluster_overl.pdf} we give the fraction of clusters in a given sample that share their central galaxy (in the same projection) with a cluster in a comparison sample and pass the richness filter in both. We see that most clusters are indeed duplicated. Those that are not, fail because in one of the two samples either a more massive potential member is included or the richness falls outside the allowed range. Such differences are a first indication of sensitivity to projection effects, an issue that is discussed further in subsection \ref{sec:clus_subfind}.
    
    \begin{table}
      \caption{The fractional overlap between different cluster samples.}
      \centering
      \label{tab:cluster_overl.pdf}
      \begin{tabular}{rccc}
        \hline
        \multirow{2}{*}{Base sample} & \multicolumn{3}{c}{Comparison sample}\\
        \cline{2-4}
        & CS60 & CS120 & CS250\\
        \hline
        CS60 & 1.0 & 0.876 & 0.736\\
        CS120 & 0.874 & 1.0 & 0.783\\
        CS250 & 0.758 & 0.808 & 1.0\\
        \hline
      \end{tabular}
    \end{table}
Notice that the algorithm described above allows a given galaxy to be considered a member of more than one cluster. Although the majority of our simulated clusters do not have such overlaps, they are not negligible; the fraction of clusters which share at least one galaxy
with another cluster in the same projection is 18.8, 21.8 and 26.7 per cent for CS60, CS120 and CS250, respectively. The average number of galaxies in these overlaps is $\sim 14$, which should be compared with the mean number of galaxies per cluster which is 37 to 46. In order to check the importance of the overlaps, we have repeated our analysis using only the $\sim 80\text{--}75$ per cent of clusters which have no overlap. These are clearly a biased subset with respect to their surroundings, and as a result the stacked profiles change noticeably.  However, the conclusions we draw below are not significantly affected, and for the rest of this paper we show only results based on the full cluster samples, noting
that the redMaPPer algorithm also allows a given red galaxy to be considered part of more than one cluster, albeit by assigning probabilities to each potential membership based on the galaxy's photometric redshift, its projected separation from each cluster centre, and the richness of the clusters. The consistent use of such probabilities is the principal difference between the actual redMaPPer algorithm and the simplified version we use here.

    \subsubsection{Cluster concentrations}\label{sec:cgal}
    
    At the core of the following analysis is the separation of each cluster sample into two equal subsamples with identical richness distributions, but disjoint distributions of concentration $c_{\rm gal}$ as introduced by \cite{miyatake_evidence_2016}. This concentration is based on the mean projected distance from cluster
centre of red galaxy members, $c_{\rm gal} = R_c/\langle R_{\rm mem}\rangle$ where in our case 
    \begin{equation}
    \left<R_{\mathrm{mem}}\right> = \frac{1}{N_{\mathrm{mem}}}\sum\limits_i^{N_\mathrm{mem}} R_{\mathrm{mem},i} .
    \end{equation}
 We classify a particular cluster as high or low concentration, depending on whether $c_{\rm gal}$ lies above or below the median for all
 clusters of the same richness. For richness values with fewer than 200 clusters in a given sample, we bin together neighbouring richness bins to exceed this number before determining the median. For the observed clusters \cite{miyatake_evidence_2016} binned clusters by both richness and redshift before determining the median, but redshift binning is not necessary for the simulated samples since they are all taken from the same simulation output.
    
    \subsubsection{The cluster-halo correspondence}\label{sec:clus_subfind}
    
It is not straightforward to connect a galaxy cluster defined in projection with a specific three-dimensional cluster, in our case a specific \textsc{subfind} halo. The idealised model of a spherically
symmetric cluster centred on its most massive galaxy and containing all the cluster members identified in projection corresponds poorly to most of the clusters identified either in the simulation or, most likely, in the SDSS. In almost all cases, the galaxies identified as members in 2D reside in multiple 3D objects distributed along the line-of-sight. This makes the cross-identification between 2D and 3D ambiguous.
    
Here we consider two possibilities for defining the 3D counterpart
of each 2D cluster: the dark matter halo that hosts the central galaxy and the one that hosts the largest number of member galaxies. The former definition follows the logic of the cluster centring, while the latter ensures that the richness of the 3D counterpart corresponds most closely to that of the 2D system. It is interesting to see how often these definitions coincide, i.e., how often the central galaxy is actually part of the dominant galaxy aggregation along the line-of-sight. We give in \autoref{tab:pairing_stats} the fraction of clusters in each of our three samples for which both methods lead to the same FoF halo. These numbers show that that the two definitions are generally in good agreement, and that this is better for smaller maximal depth offsets
and for more concentrated clusters. These trends reflect the projection effects discussed in detail in \autoref{sec:projection_effects}.
   
It is also interesting to see how many of the potential cluster members identified in 2D are, in fact, part of the same 3D object. For each of our clusters we find the maximal fraction of its members contained in a single 3D FoF halo. The third column of \autoref{tab:pairing_stats} then gives the average of these fractions. This can be compared with the average fraction
of its members contained in the FoF halo of its central galaxy (fourth column) and with the average expected as a result of our background correction,
$\langle\lambda/N_g\rangle$, given in the last column.

The values for $\langle F_{\rm biggest}\rangle$, $\langle F_{\rm central}\rangle$ and $\langle\lambda/N_g\rangle$ in \autoref{tab:pairing_stats} show that we consistently find more 'foreign' galaxies in our clusters than we would expect from contamination by a uniform background. The more concentrated clusters have contamination ratios close to, yet still a few percent below the expected ones. The low-concentration clusters have contamination fractions more than twice the expected values. We therefore conclude that the identified clusters are biased towards arrangements of multiple objects along the LoS, especially in the \lcg case. Again, this is very much in line with our discussion on the preferential selection of aligned systems in \autoref{sec:projection_effects}.

    \begin{table}
      \caption{The fraction of clusters where the central galaxy resides in the FoF halo contributing the largest number of potential 2D members; the mean fraction of such members in this halo; the mean fraction of such members in the FoF halo of the central galaxy; the mean membership fraction predicted by 'standard'  background subtraction.}
      \centering
      \label{tab:pairing_stats}
      \begin{tabular}{rrcccc}
        \hline
        \multirow{1}{0.65cm}{Subs.} & \multirow{1}{*}{Sample} & $F_{\rm centred}$ & $\langle F_{\rm biggest}\rangle$ & $\langle F_{\rm central}\rangle$ & $\langle\lambda/N_g\rangle$\\
        \hline
        \multirow{3}{0.65cm}{All} & CS60 & 0.93 &  0.826 & 0.803 & 0.922 \\
        & CS120 & 0.903 & 0.755 & 0.726 & 0.856 \\
        & CS250 & 0.848 & 0.635 & 0.595 & 0.743 \\
        \hline
        \multirow{3}{0.65cm}{high $c_{gal}$} & CS60 & 0.983 & 0.880 &  0.874 & 0.922 \\
        & CS120 & 0.973 & 0.819 & 0.812 & 0.855 \\
        & CS250 & 0.948 & 0.709 & 0.697 & 0.742 \\
        \hline
        \multirow{3}{0.65cm}{low $c_{gal}$} & CS60 & 0.876 & 0.772 & 0.732 & 0.923 \\
        & CS120 & 0.833 & 0.69 &  0.64 &  0.857 \\
        & CS250 & 0.749 & 0.561 & 0.494 & 0.744 \\
        \hline
      \end{tabular}
    \end{table}
    
\section{Results In Projection}\label{sec:projection}

We are now in a position to investigate whether the assembly bias
and splashback features identified in SDSS data  by \cite{miyatake_evidence_2016} and \cite{more_detection_2016} are reproduced when our simplified version of the redMaPPer algorithm is applied to the public Millennium Simulation data . We begin by comparing the observed mean galaxy and mass profiles to directly analogous profiles for CS250, finding that both the surprisingly strong assembly bias and the unexpected properties of the apparent splashback signal are reproduced well. Most differences can be ascribed to the finite size of the simulation or to the simplifications of our cluster identification scheme. We then use our three cluster catalogues to investigate the dependence of these successes on $\Delta z_m$, the maximal depth offset allowed for potential cluster members, finding that the assembly bias signal is sensitive to this parameter but the splashback features are not. Finally we look in more detail at the radial dependence of the ratio of the profiles of low- and high-concentration clusters. Later sections employ the full 3D information available for the simulation to explore the origin of the observed features, and vary our cluster identification scheme to demonstrate how its imprint on the measured profiles can confuse identification of the splashback signal. 

\subsection{Comparison of profiles for SDSS and CS250}

\begin{figure}
      \centering
      \includegraphics{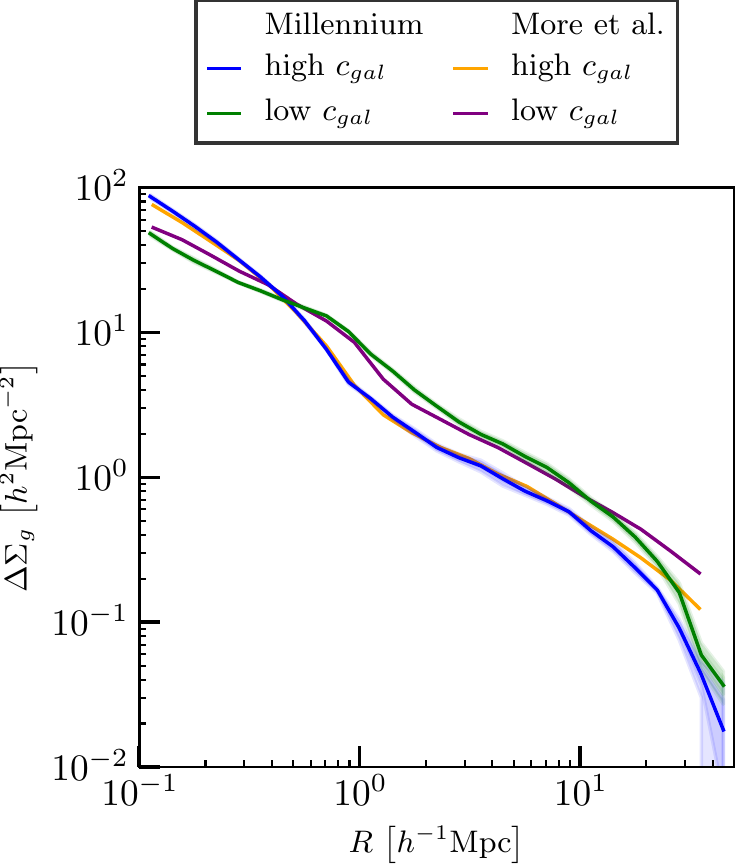}
      \caption{Mean surface number density profiles $\Delta\Sigma_g$ for galaxies with $M_{i} < -20.20$ surrounding clusters in the low- and high-concentration subsamples of CS250 are compared with observational results from \protect\cite{more_detection_2016}.}\label{fig:deep_gprof}
    \end{figure}

\begin{figure}
      \centering
        \includegraphics{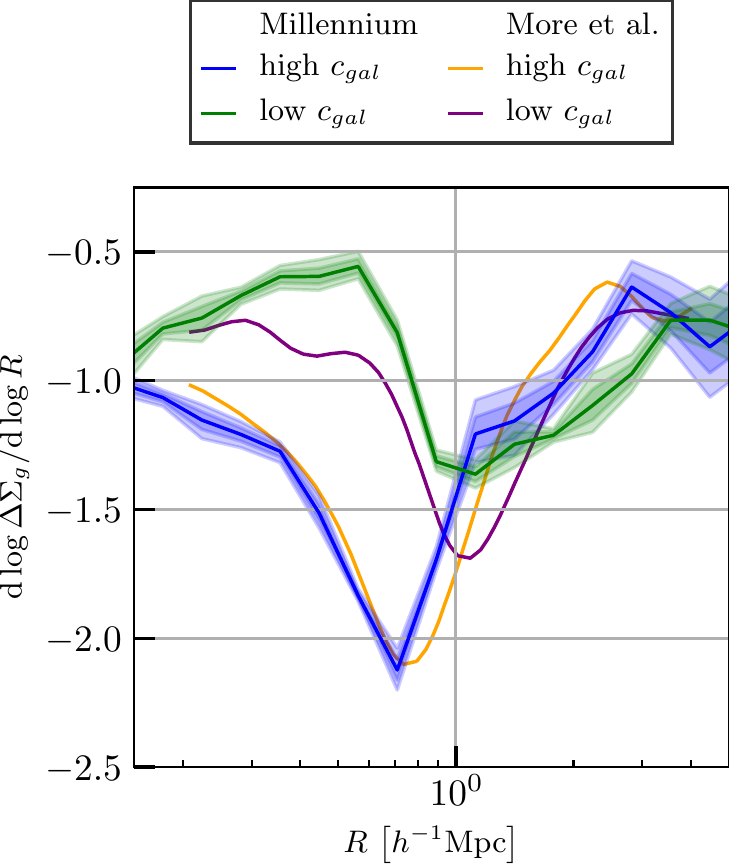}
      \caption{The logarithmic derivatives of the $\Delta\Sigma_g$ profiles for CS250 shown in \protect\autoref{fig:deep_gprof} are compared with those plotted for their SDSS clusters by More et al. (2016).}\label{fig:deep_gderiv}
    \end{figure}

\begin{figure}
      \centering
      \includegraphics{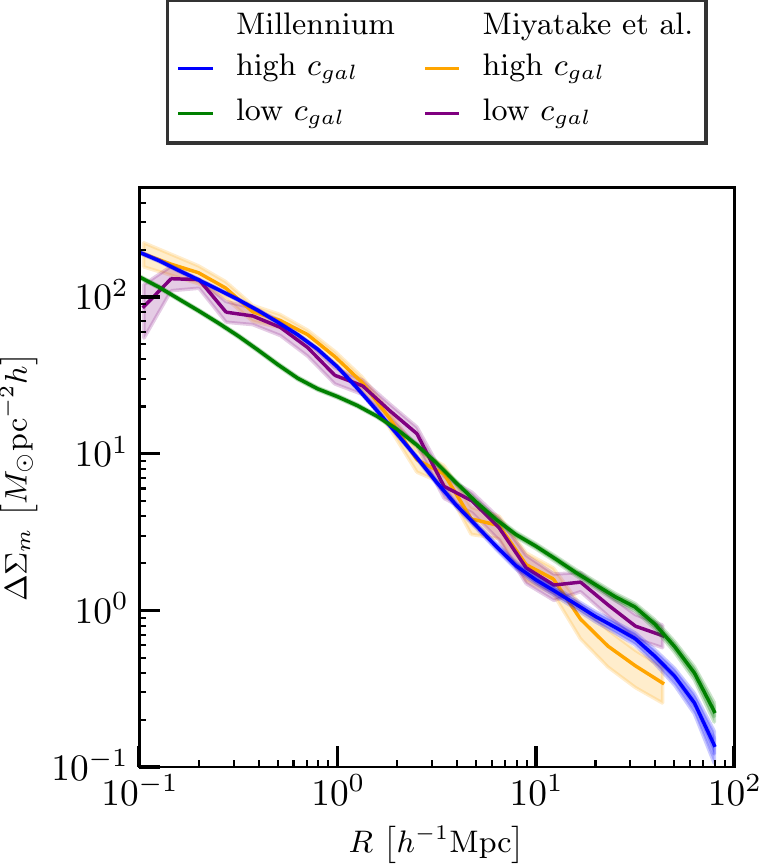}
      \caption{The mean lensing observable \dsm for high- and low-concentration clusters in the CS250 sample is compared to observational results for SDSS clusters from \protect\cite{miyatake_evidence_2016}.}\label{fig:deep_mprof}
    \end{figure}

We collect the main profile results for the CS250 sample in Figures \ref{fig:deep_gprof} to \ref{fig:deep_mprof}. Here and in the following, unless noted otherwise, the solid line represents the median value from 10000 bootstrap resamplings of the relevant cluster sample. The shaded regions denote the 68 per cent (darker) and 95 per cent (lighter) confidence intervals around this median.
    
We calculate the mean galaxy surface number density profile for each cluster sample as 
\begin{equation}
          \Delta\Sigma_g(R)=\Sigma_g(R)-\bar\Sigma_g \label{eqn:dsg}
    \end{equation}
where we use all galaxies brighter than $M_{i} = -19.43 + 5\log_{10}h$, not just the red ones, and we impose no maximal depth offset from the
cluster. $\Sigma_g(R)$ is then the mean over all clusters of the surface number density of such galaxies in an annular bin at projected distance $R$ from the central galaxy, and $\bar\Sigma_g $ is the mean surface density over the full projected area of the simulation.

\autoref{fig:deep_gprof} shows that CS250 reproduces the findings of \cite{more_detection_2016} remarkably  well. The deviation at large scales ($>20\si{\Mpch}$) is expected and reflects a lack of large-scale power in the Millennium Simulation due to its finite size. The offset between the high- and low-concentration subsamples at $R>3\si{\Mpch}$ shows that the simulation reproduces the strong assembly bias seen in the SDSS data. On small scales ($< 300\si{\kpch}$) the number density profile is slightly too steep for the high-concentration clusters, but shows otherwise very good agreement, while there is an offset of $0.1\text{ dex}$ for the low-concentration subsample inside $\SI{400}{\kpch}$. The most notable differences are on intermediate scales, especially in the range $\SI{1}{\Mpch}\leq R \leq \SI{3}{\Mpch}$ for the low-concentration case. For high-concentration clusters the agreement in this range is excellent and extends out to well beyond \SI{10}{\Mpch}. This is the radial range where splashback features are expected, but is also comparable to the radius, $R_c$, used operationally to define clusters. These differences are highlighted in the radial variations of the profile slope, which we look at next. 
        
    In \autoref{fig:deep_gderiv} we plot the logarithmic derivative $\mathrm{d}\log\Delta\Sigma_g/\mathrm{d}\log R$ over a restricted radial range for these same two CS250 subsamples, comparing with the same quantity for SDSS clusters as plotted by \cite{more_detection_2016}. The simulated curves appear noisier than those observed This is at least in part because of the more direct derivative estimator used here. Nevertheless, we reproduce the main features highlighted by \cite{more_detection_2016}, who identified the position of the minimum of these curves (i.e. the steepest profile slope) as their estimate of the splashback radius. The minima occur at similar radii in the observed and simulated data which, as \cite{more_detection_2016} pointed out, are smaller than expected given lensing estimates of cluster mass. Further the minimum is deeper for the high concentration sample and occurs at smaller radius, whereas the opposite is expected from earlier work on the dependence of splashback radius on halo accretion history (and hence concentration, see \cite{diemer_dependence_2014}). In addition, there are clear differences between the observed and simulated curves. In particular, the profiles of simulated low-concentration clusters are clearly shallower than observed in the range 200$\si{\kpch}$ $<R<\SI{1.5}{\Mpch}$.

    We discuss these features in more detail in \autoref{sec:rc_influence}, showing them to result from the superposition of effects induced by the cluster selection algorithms on the true splashback signal.
    
Mean mass density profiles can be computed much more straightforwardly for our simulated cluster samples than is possible observationally, where such profiles are obtained from the correlated orientation of background galaxies induced by gravitational lensing. In order to compare with the lensing results in \cite{miyatake_evidence_2016}, we bin up the projected mass distribution of the simulation around cluster central galaxies in exact analogy to the above procedure for creating galaxy number density profiles, and we manipulate the resulting profiles to obtain the directly observable quantity, 
    \begin{equation}
      \dsm(<R) = \Sigma_m(<R) -\Sigma_m(R) . \label{eqn:dsm}
    \end{equation}
Here, $\Sigma_m(R)$ is the surface mass density profile analogous to
$\Delta\Sigma_g(R)$  above, while  $\Sigma_m(<R)$ is the mean of this quantity over projected radii interior to $R$.  Note that despite the similarity in notation (which we have inherited from earlier work)
$\dsm(<R)$ is not directly analogous to $\Delta\Sigma_g(R)$ and will differ from it in shape even if the projected mass and galaxy number density profiles parallel each other exactly. 
    
In \autoref{fig:deep_mprof} we compare $\dsm(<R)$ obtained in this way for the high- and low-concentration subsamples of CS250 to the profiles inferred by \citeauthor{miyatake_evidence_2016} from their SDSS lensing data. Whereas the observational data show 
at most small differences between the high- and low-concentration 
subsamples for $R < 10\si{\Mpch}$, our simulated profiles differ significantly in a way which is related to the differences seen in 
\autoref{fig:deep_gprof}. Indeed, we have plotted the surface mass density profiles $\Sigma_m(R)$ directly, and find they are very similar in shape and relative amplitude to the simulated galaxy surface density profiles of \autoref{fig:deep_gprof}. We note that the disagreement between simulation and observation is limited to low-concentration clusters -- agreement is very good for the high-concentration systems on all scales below about 15$\si{\Mpch}$.
We have found no explanation for this discrepancy. The uncertainties on the $\dsm(<R)$ inferred from lensing data are much larger than the purely statistical uncertainty in the simulation results, but below 1$\si{\Mpch}$
the simulation results for low-concentration clusters lie systematically below the observations, while beyond 3$\si{\Mpch}$ they tend to lie above
them. (Note that the coloured bands in \autoref{fig:deep_mprof} show the estimated $1\sigma$ uncertainties in the observations.) This disagreement is in line with the stronger differences between the projected galaxy profiles for the low-concentration subsample. Our findings for the differences in the inner part are close to the findings of \cite{dvornik_kids_2017} who recently investigated the mass profiles of galaxy groups. These less massive objects were identified with a different group finder (based on the FoF algorithm), but the same $c_{gal}$ projected concentration measure was used to divide the sample. While they found a similar split at small scales in the lensing profiles, they did not see a significant signal of assembly bias on large scales. This is expected around the masses of groups when splitting by concentration. 

\cite{miyatake_evidence_2016} inferred almost equal mean total masses, 
$M_{200m} \sim 2\times 10^{14}h^{-1}M_\odot$, for high- low-concentration clusters from their measured $\dsm(<R)$ profiles.  Processed in the same way, our simulated profile for high-concentration clusters would give a very similar answer, whereas that for low-concentration clusters would give a lower value by a few tens of percent. (For $M_{200m} = 2\times 10^{14}h^{-1}M_\odot$, $R_{200m} = 1.5 \si{\Mpch}$, in the middle of the range where simulated and observed $\dsm(<R)$ agree best.) Thus the overall mass-scale of the clusters identified in the \cite{guo_dwarf_2011} galaxy catalogues by our redMaPPer-like algorithm is close to that of the SDSS clusters studied by \cite{miyatake_evidence_2016} and \cite{more_detection_2016}.

    \subsection{The influence of cluster selection depth}
    
    The simulation results shown in the last section all referred to CS250 
for which any red galaxy projected within $R_c$ is considered a potential cluster member, regardless of its distance in depth (``redshift'') from the central galaxy. As noted previously, \cite{rykoff_redmapper_2014} estimate the $1\sigma$ uncertainty in the photometric redshift of an individual red SDSS member galaxy to increase with distance and to be $90h^{-1}$ Mpc at $z=0.25$, the median redshift of the SDSS cluster sample. Thus many of the observed clusters may be better localised in depth than in the CS250 catalogue. In this section we compare galaxy and mass profiles among our three simulated cluster catalogues, CS250, CS120 and CS60, for which the depth selection parameter $\Delta z_m = 250, 120$ and 60$\si{\Mpch}$, respectively. This allows us to assess how strongly the effective selection depth of clusters affects their apparent splashback and assembly bias signals. We find that effects are small on the former, but can be substantial on the latter.

\begin{figure*}
        \begin{minipage}{.475\textwidth}
          \centering
          \includegraphics{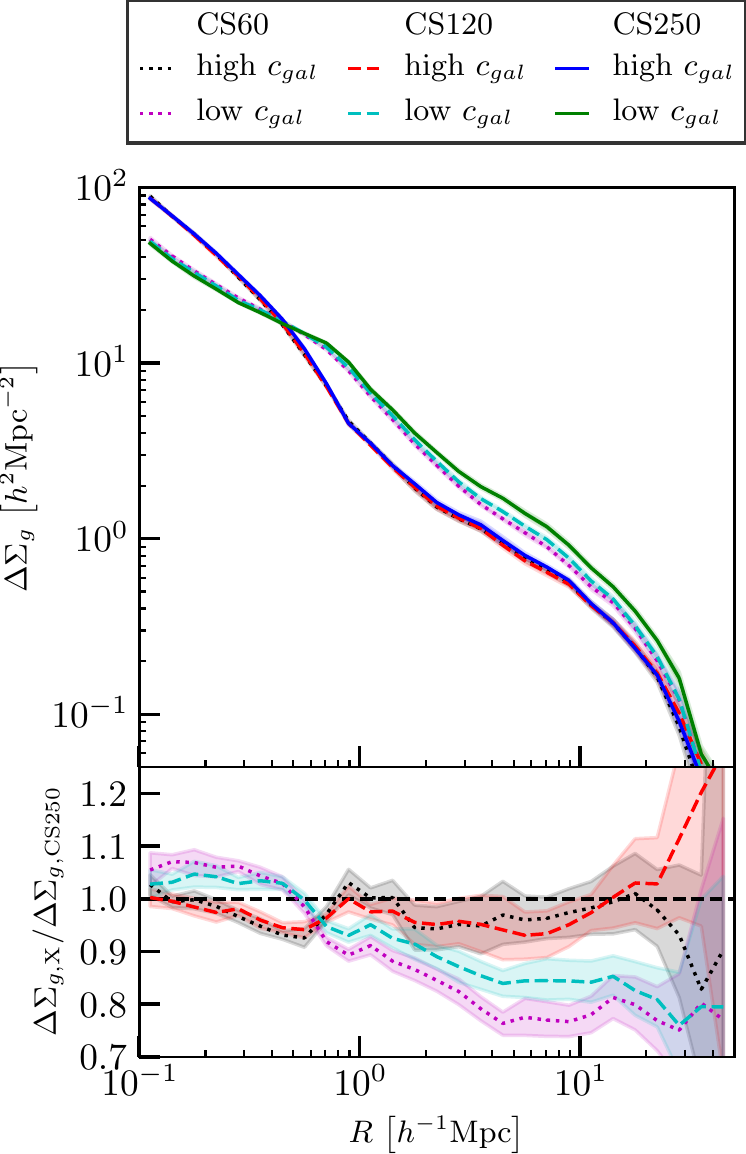}
          \caption{Comparison of the \dsg profiles for the \hcg and \lcg subsamples of our three simulated cluster catalogues (upper panel) and ratios of the profile amplitudes for CS120 and CS60 to that for CS250 (lower panel).}\label{fig:gprof_rats}
        \end{minipage}
        \hspace{0.04\textwidth}
        \begin{minipage}{.475\textwidth}
          \centering
          \includegraphics{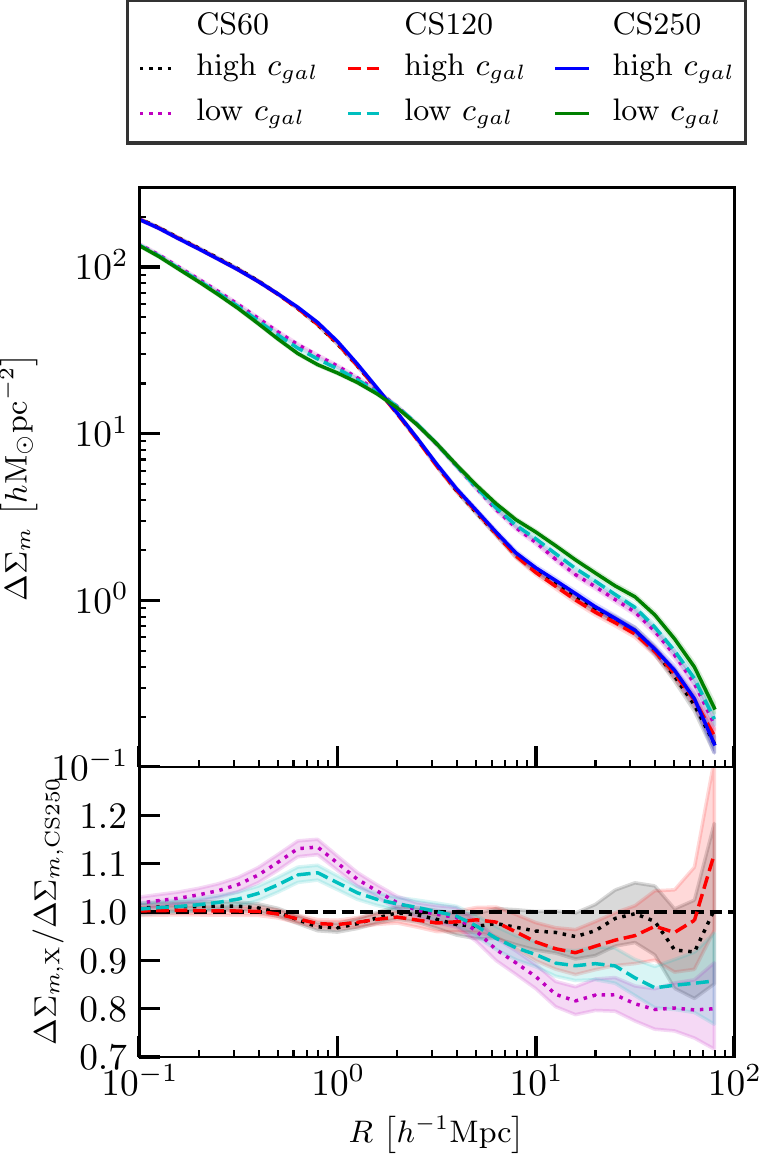}
          \caption{Comparison of the \dsm profiles for the \hcg and \lcg  subsamples of our three simulated cluster catalogues (upper panel) and ratios of the profile amplitudes for CS120 and CS60 to that for CS250 (lower panel).}\label{fig:mprof_rats}
        \end{minipage}
        \end{figure*}

\autoref{fig:gprof_rats} shows the overall shape of the projected galaxy number density profiles to be very similar in the three cluster catalogues. The high concentration profiles differ from each other by at most 10 per cent within $R_c$ and remain within the same bound out to $\sim 20 \si{\Mpch}$. Beyond this point the uncertainties increase drastically and the ratios of the profiles with smaller $\Delta z_m$ quickly depart from unity but stay within a less than the 68-percentile of the bootstrap distribution of it. The variation is somewhat smaller for low-concentration clusters and is also below 10 per cent within $R_c $, but also below 25 per cent all the way out $\sim 30 \si{\Mpch}$. Beyond $R_c$ the profile amplitude of low-concentration clusters decreases with decreasing $\Delta z_m$ at all separations where it is reliably determined.

This level of agreement is such that all three catalogues agree almost equally well with observation. In the profiles themselves, systematic differences only start to become noticeable outside $R_c$ and the largest effect is the shift in the large-scale amplitude of the profile for the low-concentration clusters, which, as we will see below (in  \autoref{ssec:bias_ratios_2d}) is enough to affect the apparent level of assembly bias significantly. At the intermediate radii relevant for splashback detection, the profile shapes are sufficiently similar that curves like those of \autoref{fig:deep_gderiv} show almost no dependence on  $\Delta z_m$.        

The \dsm profiles (shown in \autoref{fig:mprof_rats}) also vary only slightly as a function of effective cluster depth, $\Delta z_m$, with shifts of similar amplitude to those seen in the projected galaxy number density profiles. For high-concentration clusters these are even smaller than for the previous case, while for low-concentration clusters they are larger within $R_c$  and have the effect of increasingly smoothing the sudden changes in slope seen in the CS250 profile as $\Delta z_m$ decreases. For both cases the amplitude of the profiles on large scales is decreased
for smaller $\Delta z_m$, though by less than 25 per cent out to $\sim 50 \si{\Mpch}$.

    \subsection{Profile ratios and assembly bias}\label{ssec:bias_ratios_2d}
    
\begin{figure*}
                \hspace*{-0.02\textwidth}
                \begin{tabular}{p{.475\textwidth} p{0.00\textwidth} p{.475\textwidth}}
                \includegraphics{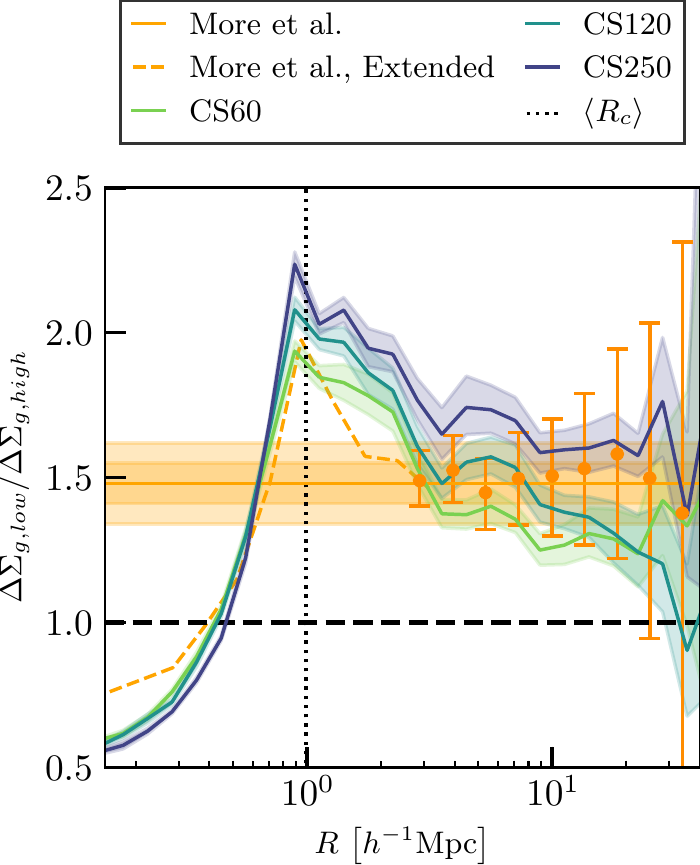}
        \caption[]{The ratio of the projected galaxy number density profiles of the \lcg and \hcg subsamples of our three simulated cluster catalogues (solid lines surrounded by their 68 per cent confidence regions). Points with error bars are observational data taken directly from \protect\cite{more_detection_2016}, while the continuation of these data to smaller scales (the dashed orange line) was calculated from the individual profiles in their paper. The dotted vertical line indicates $\langle R_c\rangle$ for the simulated clusters. The horizontal orange band is the observed assembly bias signal quoted by \protect\cite{more_detection_2016} with its 68 and 95 per cent confidence ranges.}\label{fig:surf_dens_rat_comp} & &
                \includegraphics{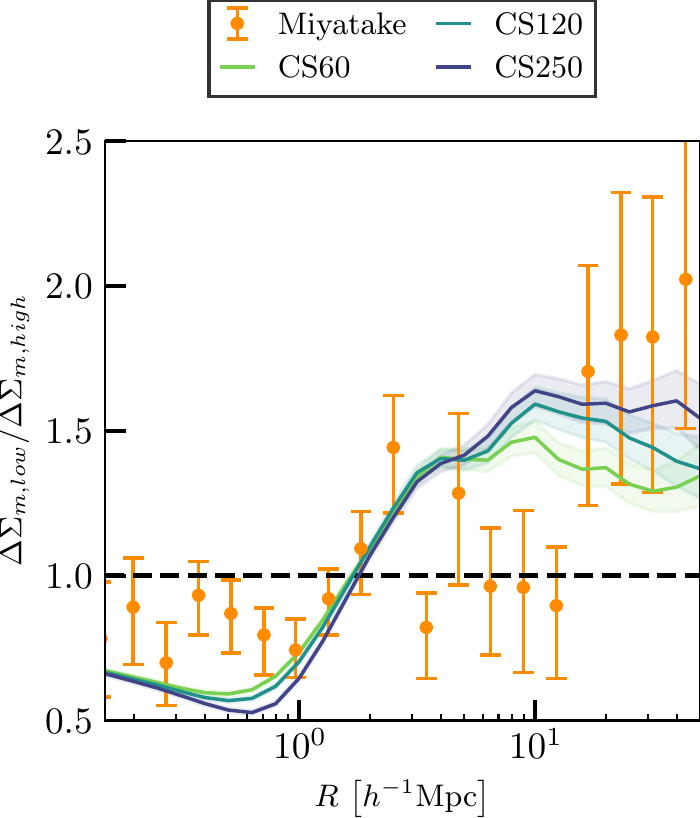}
       \caption[]{Ratios of \dsm for the high- and low-concentration subsamples of our three cluster catalogues (solid lines with their 68 per cent confidence ranges). Points with error bars are results derived from the gravitational lensing signal of SDSS clusters by \protect\cite{miyatake_evidence_2016}.}\label{fig:mat_rat_cgal}
                \end{tabular}
    \end{figure*}

By taking the ratio of the profiles discussed in the previous section we can obtain a measure of the relative bias of high- and low-concentration clusters at fixed cluster richness, hence of {\it assembly bias}. In \autoref{fig:surf_dens_rat_comp} we show this ratio for the $\Delta\Sigma_g$ profiles as a function of projected separation for our three catalogues of simulated clusters. In order to measure the large-scale bias, \cite{more_detection_2016} only plotted this ratio at $R \geq \SI{3}{\Mpch}$ (the orange points with error bars in \autoref{fig:surf_dens_rat_comp}). However, since they give the individual profiles for high- and low-concentration clusters, it is straightforward to reconstruct the observed ratio on smaller scale. We show this as a dashed orange line in \autoref{fig:surf_dens_rat_comp}.
      
The observed and the simulated ratios show similar behaviour which separates into three distinct radial regimes. At $R \geq \SI{3}{\Mpch}$, the relative bias varies little and the observed value of $1.48 \pm 0.07$ matches very well that for CS250 outside of $R=\SI{8}{\Mpch}$. CS120 gives a somewhat smaller value fitting the observations well between 3 and $\SI{10}{\Mpch}$, while at larger $R$ it is still within about $1\sigma$. CS60 has even weaker relative bias barely within $1\sigma$. Both these signals appear to decline with increasing $R$. The behaviour at smaller scales differs markedly on either side of a sharp peak which, for the simulated clusters, occurs almost exactly at $\langle R_c\rangle \sim \SI{1}{\Mpch}$, coinciding with that for the observed clusters. At smaller $R$, the ratio of the profiles increases smoothly and strongly with $R$, reflecting the requirement that the two cluster subsamples have similar richness but systematically different values of $\langle R_{\rm mem}\rangle$. This also enforces a ratio substantially above unity at $R=R_c$. At intermediate radii, $R_c<R<3\si{\Mpch}$, the ratio has to decline from the high value at the peak to the more modest value characteristic of the large-scale assembly bias.
In all three samples there is a noticeable change in slope just outside $2\si{\Mpch}$ which appears to reflect true splashback effects (see \autoref{sec:3Denvironment}). 

These properties demonstrate that the operational definition of clusters has a substantial effect on the ratio of the profiles  out to at least $3\si{\Mpch}$. These effects must therefore be present also in the individual profiles, and hence must affect their use for identifying splashback features.  In addition, the variation of the ratios at large $R$ among our three cluster catalogues shows that the apparent assembly bias signal is significantly affected by projection effects. 
    
The ratio of the \dsm profiles for the high- and low concentration subsamples of each of our three simulated cluster catalogues are shown in \autoref{fig:mat_rat_cgal} in exactly analogous format to \autoref{fig:surf_dens_rat_comp}. They are compared to observational results taken directly from \cite{miyatake_evidence_2016}. The difference in shape between the simulation curves in Figures~\ref{fig:mat_rat_cgal} and~\ref{fig:surf_dens_rat_comp} is due primarily to the conversion of $\Sigma_m(R)$ to $\Delta\Sigma_m(<R)$. A ratio plot constructed using $\Sigma_m(R)$ directly is quite similar to \autoref{fig:surf_dens_rat_comp}, although the peak at $\langle R_c\rangle$ is less sharply defined. The behaviour of the observational points in \autoref{fig:mat_rat_cgal} is quite erratic and looks rather implausible when compared with the smooth variation predicted by the simulation. Over the ranges $3\si{\Mpch}<R<14\si{\Mpch}$ and $R>15\si{\Mpch}$ the predicted assembly bias signal is almost constant, but over the first range it is much larger than and apparently inconsistent with that observed, whereas over the second it is smaller than and again apparently inconsistent with that observed. It is our impression that the uncertainties of these observational points are too large for secure interpretation to be possible.

The differences in large-scale assembly bias between our three simulated cluster catalogues are similar to those seen for the cluster number density profiles of \autoref{fig:surf_dens_rat_comp}, although pushed out to systematically larger radii.  Again this is a consequence of the conversion from $\Sigma_m(R)$ to $\Delta\Sigma_m(<R)$. On small scales the simulation curves lie well below the observational points. This is a restatement of the fact that the simulated profiles in \autoref{fig:deep_mprof} differ much more at these radii than the observed profiles.

  \section{The 3D Perspective}\label{sec:three_dimensions}
    
\cite{miyatake_evidence_2016} and \cite{more_detection_2016} interpret their SDSS results under the implicit assumption that the features seen in the stacked 2D profiles correspond to similar features in the 'true' 3D profiles.
In our simulations, it is possible to test the extent to which this is the case, so in this section we compute stacked 3D profiles of mass density and of galaxy number density around the central galaxies of our three cluster catalogues, splitting them into high- and low-concentration subsamples as before using the 2D values of $c_{\rm gal} = R_c(\lambda)/\langle R_{\rm mem}\rangle$. This allows us to make plots directly analogous to those discussed above, and so to check the 2D -- 3D correspondence.  In this section all profiles are calculated in true position space rather than in redshift space. Note that we here use a standard definition of the spherically averaged mass density profile rather than some 3D analogue of \dsm. Note also that since each central galaxy can appear in one to three different projections, we give it the corresponding weight when constructing the 3D profiles in order to keep as close a correspondence as possible to the 2D results discussed previously.

\subsection{Splashback Radius}\label{sec:3d_gal_profiles}

\begin{figure*}
        \begin{minipage}{.475\textwidth}
       \includegraphics{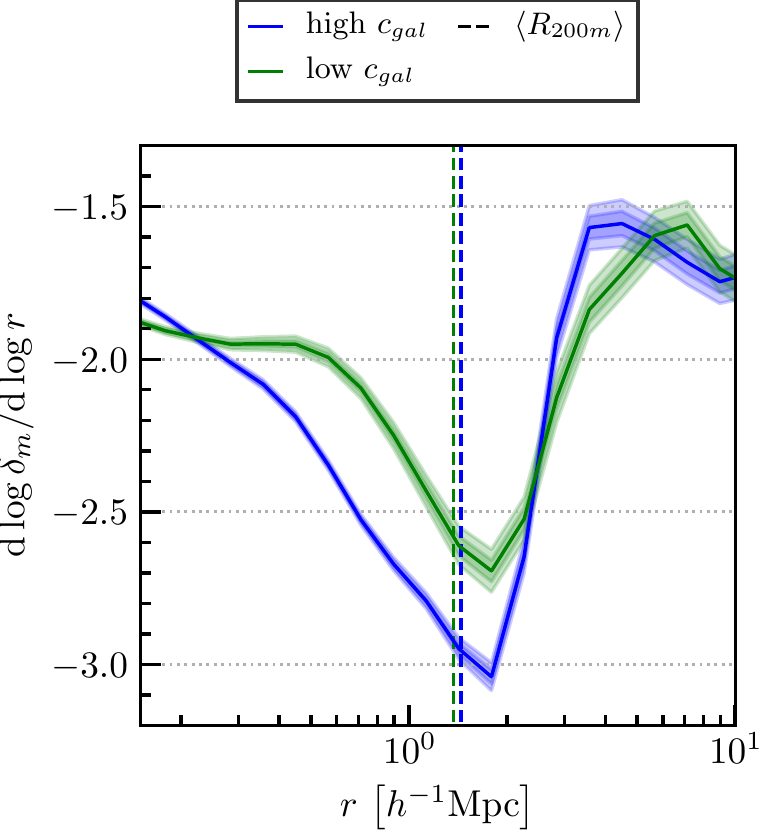}
       \caption[Log-Derivative of $\delta_m$]{Logarithmic derivative profiles of the 3D mass overdensity around the central galaxies of the high- and
low-concentration subsamples of CS250. Vertical lines mark the $R_{200m}$
values for the two samples calculated directly from their stacked mass profiles.}\label{fig:mat_dens3d_deriv_comp}
      \end{minipage}
        \hspace{0.04\textwidth}
        \begin{minipage}{.475\textwidth}
       \includegraphics{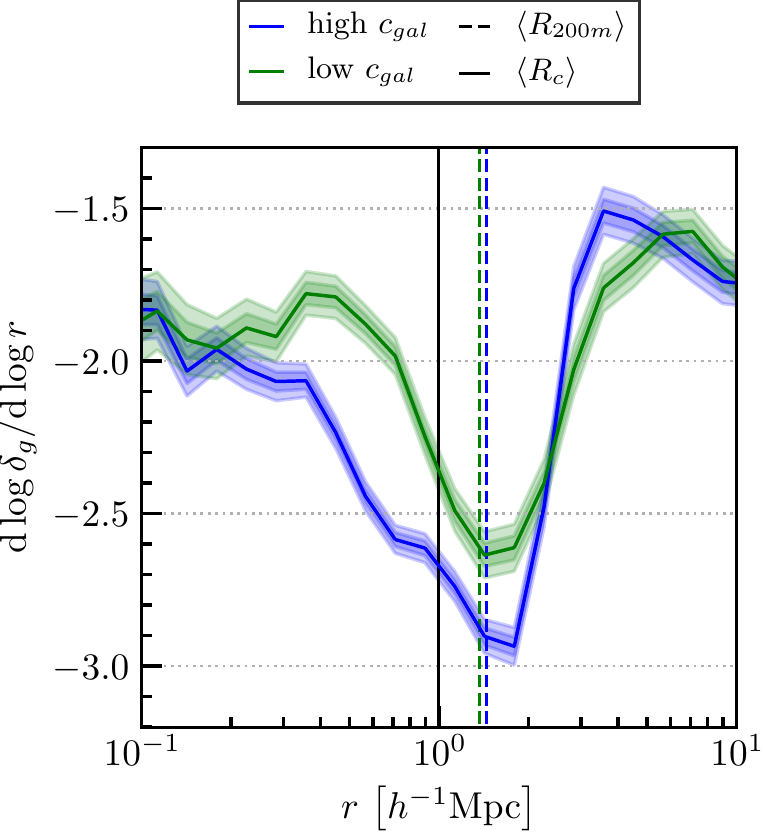}
       \caption[Log-Derivative of $\delta_g$]{Logarithmic derivative profiles of the 3D galaxy number overdensity around the central galaxies of the high- and low-concentration subsamples of CS250 in identical format to \protect\autoref{fig:mat_dens3d_deriv_comp} except that a solid vertical line indicates $\langle R_c\rangle$ for the two samples.}\label{fig:gal_dens3d_deriv_comp}
      \end{minipage}
        \end{figure*}

As was the case in 2D, we find that plots of the 3D profile slope,
analogous to those of \autoref{fig:deep_gderiv}, are very
similar for our three cluster catalogues. In Figures~\ref{fig:mat_dens3d_deriv_comp} and \ref{fig:gal_dens3d_deriv_comp} we therefore show results for CS250 only. Since recent theoretical work on splashback properties has concentrated on cluster mass profiles \citep[e.g.][hereafter DK14]{diemer_dependence_2014}, we start with a discussion of 
\autoref{fig:mat_dens3d_deriv_comp} which shows logarithmic slope (referred to as $\gamma$ below) as a function of 3D radius $r$ . 

These slope profiles show relatively smooth behaviour
with well-defined minima at $r\sim 1.8\si{\Mpch}$. The mean $M_{200m}$ values in the two sub-samples correspond to $R_{200m}\sim 1.45\si{\Mpch}$ and $R_{200m}\sim 1.37\si{\Mpch}$, so these minima occur
at $1.2 R_{200m}$ and $1.3 R_{200m}$ for the high- and low-concentration samples, respectively. These values are very close to the expected values given in \cite{more_splashback_2015} for the expected mass accretion rates at the given masses and redshift. The slopes at minimum are significantly
shallower for our stacks ($\gamma \sim -2.8$) than DK14 found for halos of similar mass ($\gamma \sim -3.5$). As shown in the Appendix, this is because such profiles depend both on
the definition of the sample to be stacked and on the details of stack
construction. In particular, DK14 scale each individual
profile to its own $R_{200m}$ and then take the median density at each
$r/R_{200m}$, whereas we take the mean density at each radius directly.
The DK14 procedure typically produces deeper and
sharper minima, hence better defined splashback radii which occur at slightly smaller radii, but it is not easily implemented on observed samples. For example, the redMaPPer samples are defined to have similar (and known) values of $R_c$ but their individual values of $R_{200m}$ are unknown. In addition, weak lensing reconstructions of the mass distribution naturally produce mean rather than median mass profiles.

The two slope profiles of \autoref{fig:mat_dens3d_deriv_comp} differ
significantly in shape.  In the inner regions ($r<R_c$) this reflects the
fact that the two samples are separated by galaxy concentration
(in practice, by $\langle R_{\rm mem}\rangle/R_c$) so that, by definition, the low-concentration clusters have shallower 2D galaxy density profiles within $R_c$ than the high-concentration clusters. \autoref{fig:gal_dens3d_deriv_comp} shows that this requirement carries over
to the 3D galaxy profiles, and it is still very visible in \autoref{fig:mat_dens3d_deriv_comp}. Similar effects are seen in Figure~14
of DK14 where they split their halo sample by
3D mass concentration. However, our results do not agree with the trend they
find for more concentrated clusters to have a shallower minimum slope and a larger splashback radius.  We have checked that if we follow their scaling and median stacking procedures, our high-concentration clusters still have a steeper minimum slope  and the same splashback radius as our low-concentration clusters. The discrepancy must reflect the difference between selecting halos by 3D mass and mass concentration and selecting clusters by 2D richness and galaxy concentration.
    
The shapes of the 3D slope profiles for the mass (\autoref{fig:mat_dens3d_deriv_comp}) and for the galaxies (\autoref{fig:gal_dens3d_deriv_comp}) are very similar, in particular,
beyond the splashback minimum. At smaller radii the features induced by
cluster selection are stronger in the galaxy profile, with a secondary minimum
just inside $\langle R_c\rangle$ which is just visible as a slight inflection in the mass profile. Overall, however, the features in the galaxy
profile are much less dramatic than in its 2D analogue, \autoref{fig:deep_gderiv}. This just reflects the fact that clusters were selected and their concentrations estimated using the 2D data

    \subsection{Large-scale environment}\label{sec:3Denvironment}
    
\begin{figure*}
                \hspace*{-0.02\textwidth}
                \begin{tabular}{p{.475\textwidth} p{0.00\textwidth} p{.475\textwidth}}
                \includegraphics{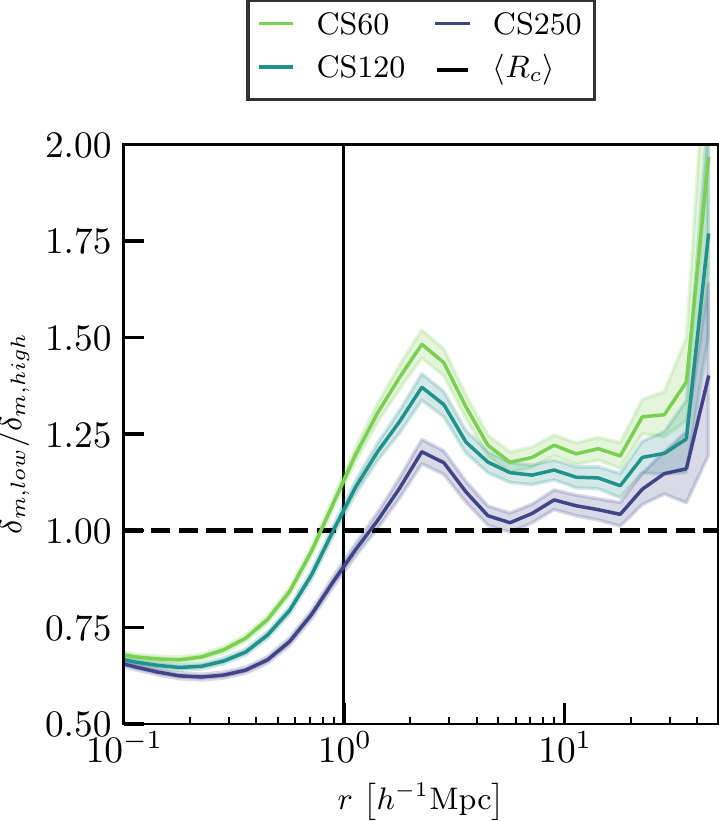}
        \caption[$\delta_{m}$ Ratios]{Ratios of the 3D mass overdensity 
        profiles of low- and high-concentration clusters for each of our
        three cluster samples. The vertical line indicates the mean cluster radius $\langle R_c\rangle$.}\label{fig:mat_dens_rat_comp} & &
                \includegraphics{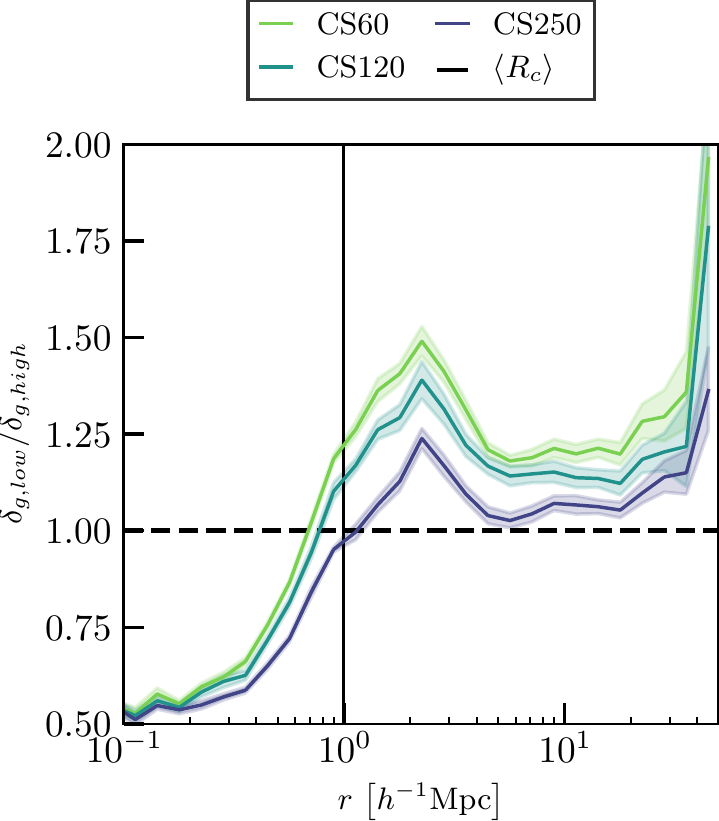}
        \caption[$\delta_{g}$ Ratios]{Ratios of the 3D galaxy number overdensity profiles of low- and high-concentration clusters for each of our three cluster samples with a vertical line indicating the mean cluster radius $\langle R_c\rangle$.}\label{fig:gal_dens_rat_comp}
                \end{tabular}
        \end{figure*}

We now look at the ratios of stacked 3D mass overdensity profiles for our low- and high-concentration clusters, and at the corresponding ratios of their galaxy number overdensity profiles. These are directly analogous to the ratios of 2D galaxy number overdensity profiles shown in \autoref{fig:surf_dens_rat_comp}. As in that figure, we here compare results for the three samples, CS60, CS120 and CS250. Ratios as a function of $r$ are shown for mass overdensities in \autoref{fig:mat_dens_rat_comp} and for galaxy number overdensities in \autoref{fig:gal_dens_rat_comp}. The shapes of the curves and their relative positions for the three samples are very similar in these two
figures.
    
In the inner regions, $r < R_c$, all curves are rapidly and smoothly rising, showing that the difference in 2D galaxy profiles resulting from our classification by concentration carries over to the 3D galaxy and mass profiles. In this regime and in both plots the ratio for CS60 is slightly larger than that for CS120 and significantly larger than that for CS250. This behaviour mirrors that of the ratio of the fractions of 2D potential members which are part of the central galaxy's FoF group (see
\autoref{tab:pairing_stats}). Interestingly, this ranking of amplitudes for the three samples persists to much larger scales and is opposite to that
seen in 2D (\autoref{fig:surf_dens_rat_comp}). Clearly, with increasing $\Delta z_m$, projection effects contribute more strongly to low- than to high-concentration clusters not only at $R\sim R_c$ but also
at much larger projected separation.

In the range $R_c < r < 5\si{\Mpch}$, all curves continue to rise to a sharp peak before dropping again to a value which remains approximately constant over the interval $5\si{\Mpch}$ $< r < 30 \si{\Mpch}$. The peak corresponds to the crossing of the derivative curves for the low- and high-concentration subsamples in Figures~\ref{fig:mat_dens3d_deriv_comp} and~\ref{fig:gal_dens3d_deriv_comp}. It thus reflects differences in the way the splashback feature merges into larger scale structure in the two cases. As noted above, it appears to be visible as a sharp change in slope in the profiles of  \autoref{fig:surf_dens_rat_comp} (see also \autoref{fig:rc_deriv_comp} below). Between $R_c$ and the peak, effects from sample definition clearly modulate galaxy overdensity profile ratios
more strongly than mass overdensity profile ratios but the difference is
quite small.

The constant profile ratios seen over the range $5\si{\Mpch}$ $< r < 30 \si{\Mpch}$ are a direct measurement of the 3D assembly bias for cluster samples split by 2D concentration. These values are significantly smaller than the 2D values inferred from \autoref{fig:surf_dens_rat_comp}. In addition, they rank in the
opposite sense with $\Delta z_m$, they are consistent between Figures~\ref{fig:mat_dens3d_deriv_comp} and~\ref{fig:gal_dens3d_deriv_comp}, and they are similar to the values expected from previous work on assembly bias for cluster mass haloes split by concentration \citep[e.g.][]{more_detection_2016}. As we will see in the next section, a clue to the origin of this difference between the 2D and 3D estimates of assembly bias comes from the largest $r$ bins in these figures where, although noisy, the ratios of the profiles rise to large values.

\section{Projection contamination}\label{sec:projection_effects}

In the preceding sections we found a number of differences in the apparent splashback and assembly bias signals between the 2D and the 3D profiles of 
our simulated galaxy clusters. These differences are present both in the mass and in the galaxy number density profiles, and they affect the low- and high-concentration subsamples to differing degrees. In this section we focus specifically on galaxy number density profiles, compiling them in the two dimensions of projected separation and line-of-sight depth so that we can compare results for the two subsamples and isolate the distribution in depth
of the galaxies which give rise to the difference in projected profiles.

Let $R$, as above, denote projected separation, and $q>0$ denote line-of-sight separation, measured either in configuration space ($q = |d|$) or in redshift
space ($q = |\pi|$). We define a set of cells of constant width in $\ln R$
and $\ln q$ and compile galaxy counts in these cells around the central
galaxies of the low- and high-concentration subsamples of each of our cluster
samples, $N_{lo}(R,q)$ and $N_{hi}(R,q)$ respectively.

\begin{figure*}
                \hspace*{-0.02\textwidth}
                \begin{tabular}{p{.475\textwidth} p{0.00\textwidth} p{.475\textwidth}}
                \includegraphics[width=.475\textwidth]{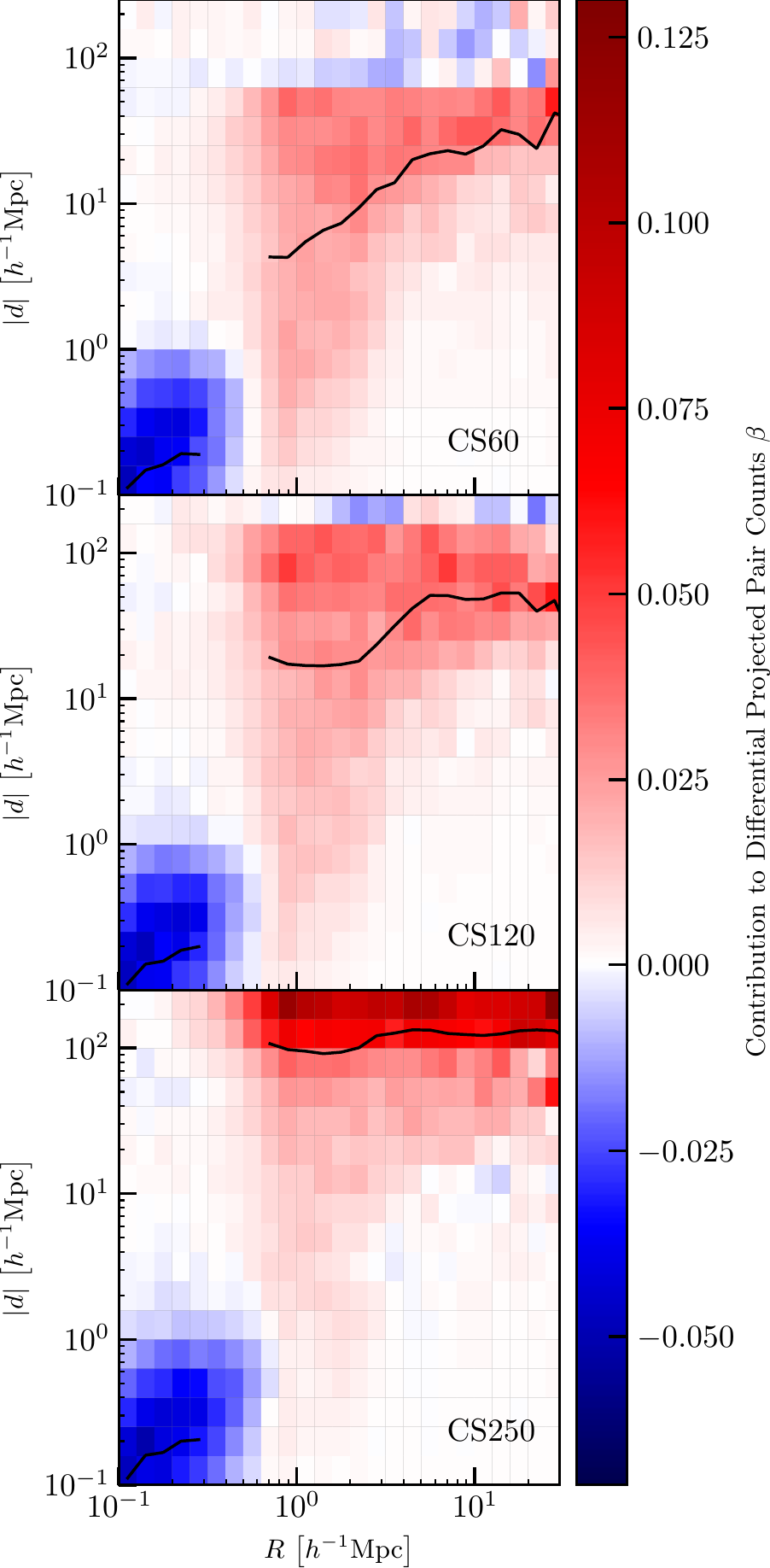}
        \caption[Differential Pair Counts]{The quantity $\beta(R,q)$ of \protect\autoref{eqn:betaRq} for the case $q= |d|$. This shows the distribution over depth $q$ of the fractional difference between the projected galaxy count profiles of the
\lcg and \hcg subsets of each of our three simulated cluster samples. The black curves give the median offset in depth of the excess counts as a function of $R$.}\label{fig:diff_pair_counts_norsd} & &
                \includegraphics[width=.475\textwidth]{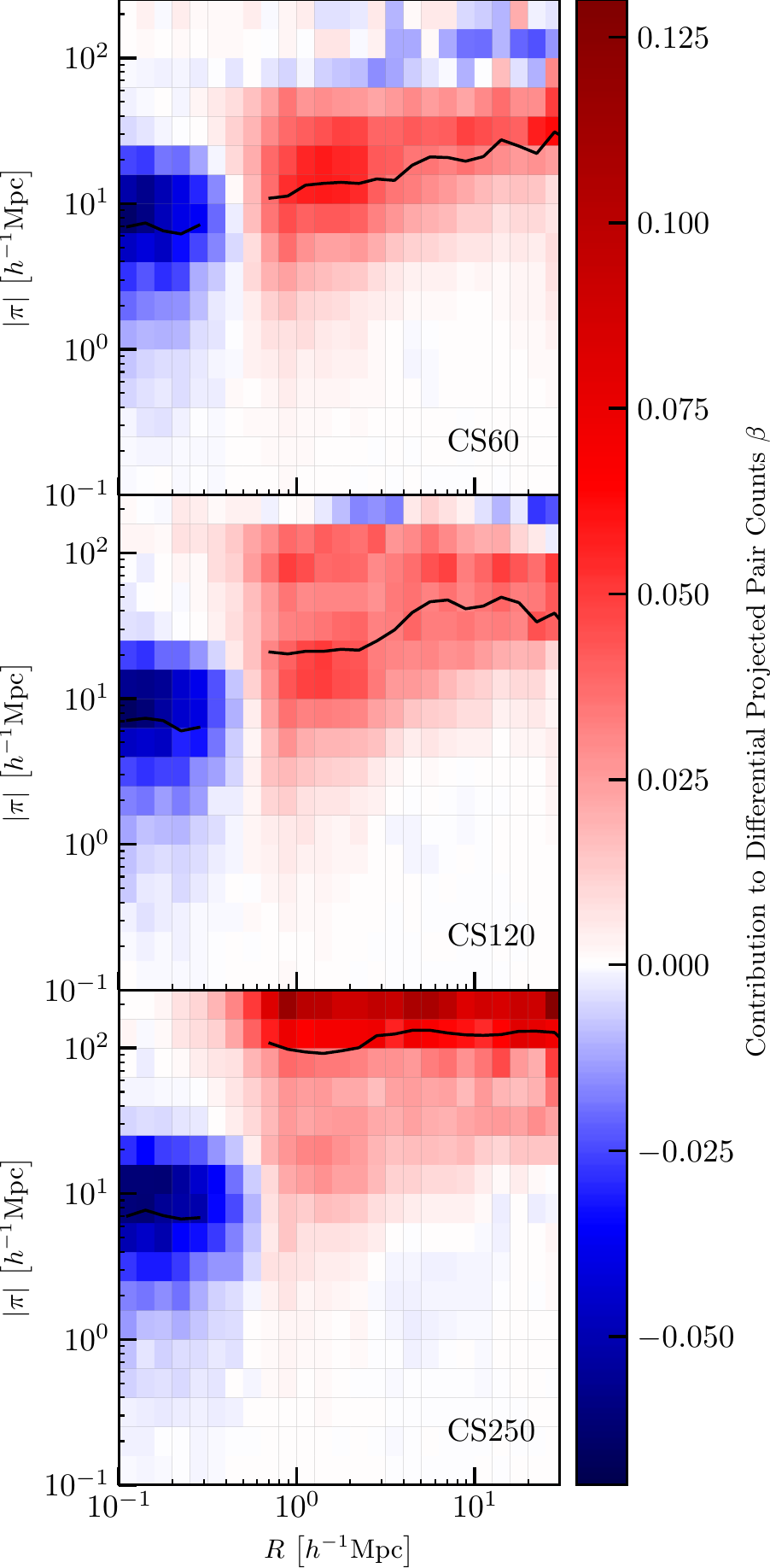}
        \caption[Differential Pair Counts]{Identical to \protect\autoref{fig:diff_pair_counts_norsd} except for the redshift space case, $q=|\pi|$.}\label{fig:diff_pair_counts} 
                \end{tabular}
        \end{figure*}

In Figures~\ref{fig:diff_pair_counts_norsd} and~\ref{fig:diff_pair_counts} we show the quantity
    \begin{equation}
     \beta(R,q) = \frac{N_{lo}(R,q)-N_{hi}(R,q)}{\sum_q [N_{lo}(R,q)+N_{hi}(R,q) - N_{c}n_{gal}V(R,q)]},\label{eqn:betaRq}
    \end{equation}
for the real-space and redshift space cases respectively. In this equation,
$N_c$ is the total number of clusters in the sample, $n_{gal}$ is the mean space density of galaxies, and $V(R,q)$ is the volume of the cell at $(R,q)$. Thus $2\sum_q\beta(R,q) = b_{lo}(R) - b_{hi}(R)$, where the assembly bias factors $b_{lo}$ and $b_{hi}$ are the ratios of the stacked 2D galaxy
number overdensity profiles of the low- and high-concentration subsamples to
that of the cluster sample as a whole. The distribution of $\beta$ over $q$
at fixed $R$ thus indicates the distribution in depth of the difference in galaxy counts which gives rise to the apparent 2D assembly bias signal.

In the inner regions ($R < 400\si{\kpch}$) the projected profile of \hcg clusters lies above that of \lcg clusters for all three samples (see \autoref{fig:surf_dens_rat_comp}). \autoref{fig:diff_pair_counts_norsd} shows that, as expected,  the additional galaxies which produce this excess lie in the inner regions of the clusters, with a median depth offset from the central galaxy 
of $150\si{\kpch}$ or less. In redshift space, the random motions within clusters move this excess out to $|\pi| \sim \SI{700}{\km\per\s}$, as shown in \autoref{fig:diff_pair_counts}. 

Beyond $R = 400\si{\kpch}$ the behaviour switches and the projected profile of \lcg clusters lies above that of \hcg clusters (again see \autoref{fig:surf_dens_rat_comp}). The galaxies which produce this excess lie
in two different ranges of depth whose relative contribution varies both with $R$ and with $\Delta z_m$. At $R < 2\si{\Mpch}$, a 'local' component centred near $R \sim |d| \sim \langle R_c\rangle$ contributes most of the excess \lcg counts in CS60, about half of them in CS120, and a minority of them in CS250, producing much of the pronounced peak seen at these $R$ in the profile ratios of \autoref{fig:surf_dens_rat_comp}. A second component, distributed relatively uniformly over $\pm \Delta z_m$, the full allowed depth for potential cluster members, contributes excess counts to the \lcg cluster profiles at all $R>R_c$ and is responsible for most of the large-scale assembly bias. It also dominates the excess counts near $\langle R_c\rangle$ in CS250. The systematic change in the relative weight of these two components with increasing $R$ results in a shift in the median depth offset of the excess counts, indicated by the black solid lines in Figures~\ref{fig:diff_pair_counts_norsd} and~\ref{fig:diff_pair_counts}. The increasing strength of the second component from CS60 to CS120 to CS250 is the cause of the increase in 2D assembly bias with $\Delta z_m$. \autoref{fig:diff_pair_counts} shows that redshift space distortions significantly smear out these two components and make them more difficult to distinguish.

These results explain why strong assembly bias is seen in 2D for CS250 and CS120 (see \autoref{fig:surf_dens_rat_comp}) but only a much weaker signal is seen in 3D
(\autoref{fig:gal_dens_rat_comp}). Many of the low-concentration clusters in these samples have significant foreground/background groups projected on their outer regions. These groups are distributed over the full depth $\pm \Delta z_m$, and are visible in Figures~\ref{fig:diff_pair_counts_norsd} and~\ref{fig:diff_pair_counts} as an excess in bins at large $q$ and $R\sim R_c$. Galaxies correlated with these foreground/background groups then produce excess galaxy counts at similar $q$ for all $R$ values shown in the plot. Since the fall-off in these counts with $R$ at the $q$ of the background group is similar to that of galaxy counts at relatively small $q$ correlated with the primary cluster, the induced apparent assembly bias is almost independent of  $R$. The rise in 3D assembly bias seen at the largest $r$ in \autoref{fig:gal_dens_rat_comp} is a result of beginning to pick up this additional correlated component in the counts around low-concentration clusters. 

The strength of this effect clearly depends on the sensitivity of the cluster identification algorithm to projection effects at $R\sim R_c$. This in turn depends both on the effective $\Delta z_m$ and on the weight assigned to potential members near the cluster edge. Hence, the apparent bias may differ between the real redMaPPer sample and our simulated samples. Nevertheless, the strong similarity seen in previous sections between the behaviour of our CS250 and CS120 samples and the SDSS sample analysed by \cite{more_detection_2016} and \cite{miyatake_evidence_2016} suggests that the assembly bias signal they found has a similar origin to that in the simulation. In the next section we will explore further the dependence of apparent splashback features on cluster definition and argue that the unexpected properties of the features detected by \cite{more_detection_2016} are a result of confusion with features imposed by the cluster selection procedure.

    \section{Cluster definition affects profile shape}\label{sec:rc_influence}
    
We have argued above that the details of our redMaPPer-like algorithm leave an
imprint on the stacked profiles of our simulated clusters. Although this is most evident in the strong peak at $R_c$ in the profile ratios of \autoref{fig:surf_dens_rat_comp} and in the steep gradient interior to this radius induced by our separation of the two subsamples by concentration, $c_{\rm gal}$, it is also visible in the crossing at $R_c$ of the individual gradient profiles of \autoref{fig:deep_gderiv} and in their minima close to and on opposite sides of this radius. In this section we investigate these effects further by varying the value of $R_c$ used to define clusters. Specifically, we set
    \begin{equation}
     R_c = 1.0 \eta \left(\frac{\lambda}{\lambda_{n}(\eta)}\right)^{0.2}\si{\Mpch}\label{eq:varlam}
    \end{equation}
and we change $\eta$. 

The variable normalisation $\lambda_{n}\left(\eta\right)$ in \autoref{eq:varlam} accounts for the fact that a given cluster will contain more galaxies within a larger 
projected radius. In the following we will consider $\eta = 2/3, 1$ (the value used in all earlier sections) and 3/2. Based on the mean galaxy number overdensity stacks of \autoref{ssec:bias_ratios_2d}, we take $\lambda_{n}\left(\eta=\frac{2}{3}\right)=74$, $\lambda_n(1) =100$, as before, and $\lambda_{n}\left(\eta=\frac{3}{2}\right)=130$. For each choice of $\eta$ we repeat the cluster selection and concentration calculation procedures of Sections~\ref{sec:clus_algo} and~\ref{sec:cgal}. Since changing $R_c$ changes the richness value $\lambda$ assigned to each cluster, we shift the richness range defining our samples ($20\leq \lambda \leq 100$ for $\eta = 1$) so that the total numbers of simulated clusters above the upper and lower limits remain unchanged. In the following we show results for $\Delta z_m = 250\si{\Mpch}$ only, since the two other cases behave very similarly.

\begin{figure}
        \includegraphics{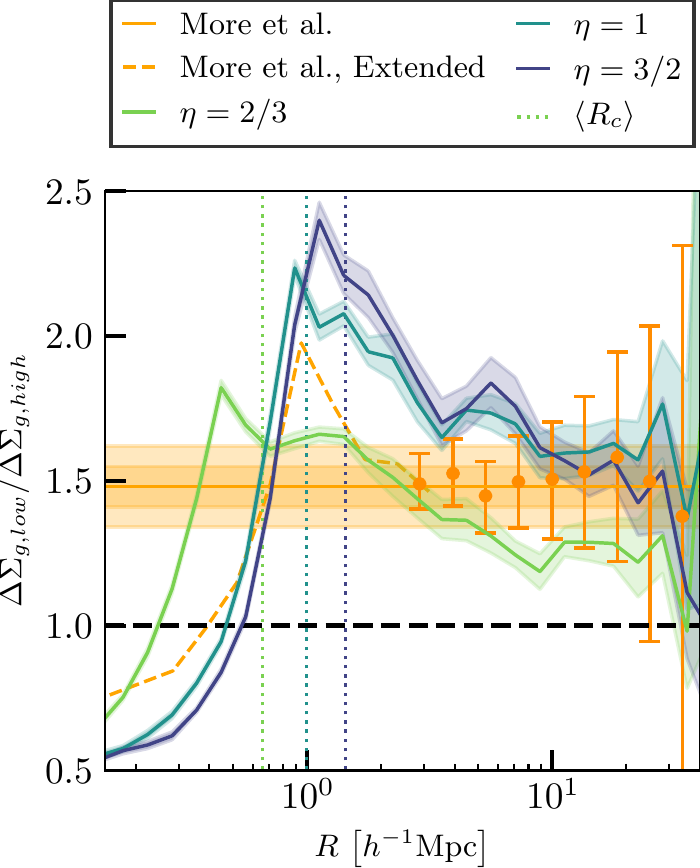}
        \caption[]{The ratio of the projected galaxy number density profiles of the \lcg and \hcg subsamples of CS250, taken from \autoref{fig:surf_dens_rat_comp}, is compared with those found for cluster samples selected with the same value of $\Delta z_m$ but with $\eta = 2/3$ and 3/2 in \autoref{eq:varlam}, rather than $\eta=1$. Points with error bars and their continuation to smaller scales are
the same as in \protect\autoref{fig:surf_dens_rat_comp}. Vertical lines indicate $\langle R_c\rangle$ for the three samples.}\label{fig:rc_bias_comp}
    \end{figure}

\autoref{fig:rc_bias_comp} repeats the observational and CS250 results from \autoref{fig:surf_dens_rat_comp} and compares them with predictions for $\eta = 2/3$ and 3/2. The peak of the profile ratio increases strongly with $\eta$ and shifts to match $\langle R_c\rangle$ in all three cases. Interestingly, the profile ratio for $\eta=2/3$ peaks at a value of 1.8 at a radius where it is 0.8 for $\eta=3/2$, and the ratio is unity for $\eta=2/3$ at a radius where it is only 0.6 for $\eta=3/2$.  Thus, changing the limiting radius defining a cluster sample not only affects its stacked profiles in their outer parts, but also close to the centre.  Beyond $R_c$, the secondary feature noted in \autoref{ssec:bias_ratios_2d} and apparently associated with true splashback effects is clearest for $\eta=2/3$ and is very weak for $\eta=3/2$. At large $R$, the strength of assembly bias increases noticeably with $\eta$. The stronger peak, the weaker splashback signal and the stronger large-scale assembly bias found with increasing $\eta$ are all consistent with the expectation that projection effects should increase in importance when clusters are identified within larger radii, hence at lower projected overdensities. Also as expected, overall the
SDSS results of \cite{more_detection_2016} behave most similarly to the $\eta=1$
curves in \autoref{fig:rc_bias_comp}. Nevertheless the large scale ratios agree equally well with the ones using $\eta=3/2$.

\begin{figure}
        \includegraphics{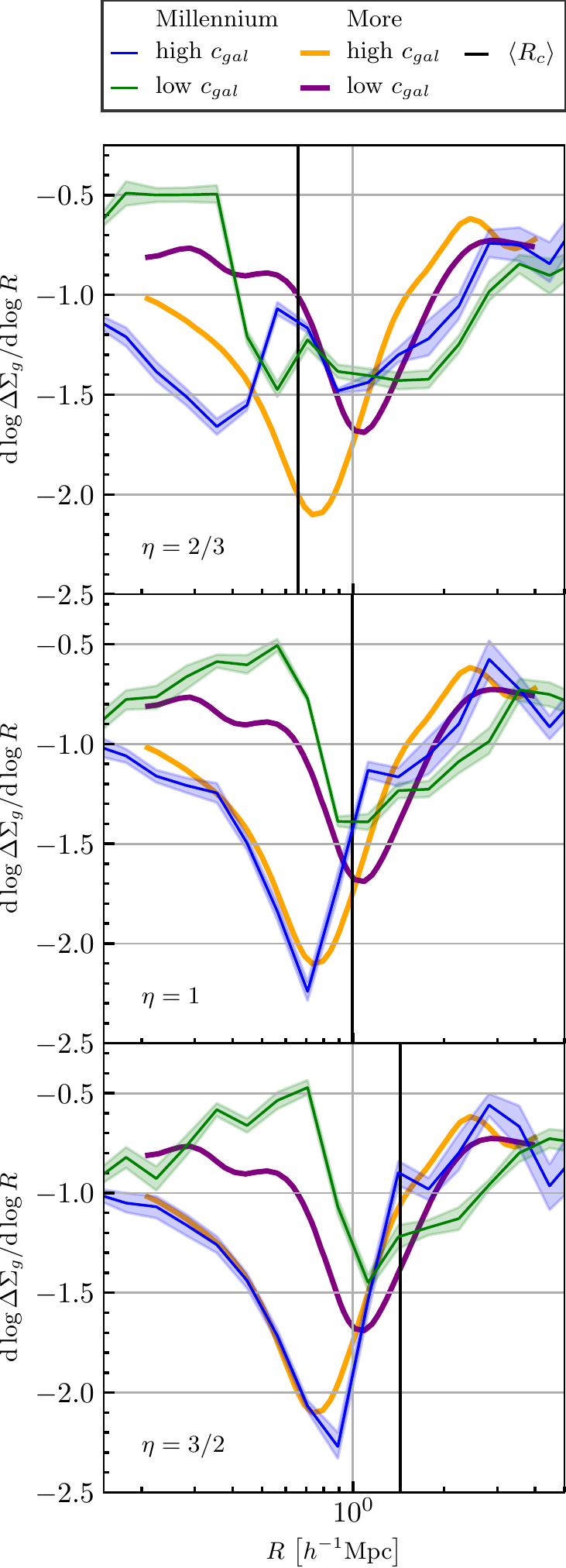}
        \caption[$\Delta\Sigma_{g}$ Radial Log-Derivaties, $R_c$ Comparison]{The logarithmic derivatives of simulated and observed $\Delta\Sigma_g$ profiles from \protect\autoref{fig:deep_gderiv} are repeated in the middle panel and compared with results from simulated cluster catalogues with the same value of $\Delta z_m$ but $\eta=2/3$ and 3/2 (top and bottom panels respectively). A solid vertical line in each panel indicates the value of $\langle R_c\rangle$ for the relevant sample.}\label{fig:rc_deriv_comp}
    \end{figure}

As shown in \autoref{fig:rc_deriv_comp}, the logarithmic derivative of $\Delta\Sigma_g$ shows a strong and complex response to $\eta$. The middle panel here is essentially a repeat of \autoref{fig:deep_gderiv}, while the upper and lower panels show similar plots for $\eta=2/3$ and $\eta=3/2$ respectively. A striking feature of these plots is that the slope profiles for the two subsamples always
cross around $R=\langle R_c\rangle$ and at a value of about -1.4. The crossing 'coincidence' is mathematically equivalent to the fact that all the profile ratios have a maximum at $R\sim R_c$ in \autoref{fig:rc_bias_comp}, which itself is easily understood as a consequence our creating subsamples with identical distributions of $\lambda$ but disjoint distributions of $c_{\rm gal}$, thus forcing the profile ratio to increase over the range $0<R<\langle R_c\rangle$. The uniform slope value at curve crossing reflects the fact that this value equals the slope for the sample as a whole, which is quite slowly varying and close to -1.4 at these projected radii.

Within the crossing point, the slope for low-concentration clusters rises rapidly to a maximum of about $\gamma=-0.5$ at $R\sim \langle R_c\rangle$, while the slope for the high-concentration clusters drops to a minimum at approximately the same radius but with a value which decreases strongly with increasing $\eta$. This behaviour is clearly a consequence of our definition of $c_{\rm gal}$ and our separation of clusters into subsamples according its value. On larger scale, the slope profiles
appear independent of $\eta$ when $R$ exceeds twice the largest value of $\langle R_c\rangle$ for the samples being compared. However, the curves for high- and
low-concentration clusters differ both from each other and from those of \cite{more_detection_2016} in this regime. In the intermediate range, $\langle R_c\rangle < R<  2\langle R_c\rangle$, the shape of the curves is set by the need
to interpolate between these two different behaviours, causing a minimum at or just outside $\langle R_c\rangle$ and a maximum at slightly larger radius in the low- and high-concentration cases respectively.

In none of these panels are the simulated curves a good fit to the observed ones.
The results for \hcg clusters match quite well for $\eta = 3/2$, but the best fit
for the \lcg clusters is for $\eta = 1$, and even here the overall depth and the general shape of the features differ significantly. Given the strong sensitivity to the cluster identification algorithm and to the splitting by $c_{\rm gal}$, it is likely that these discrepancies reflect detailed differences between the real redMaPPer and concentration definition procedures and the simplified versions used here. It is clear that it will be very difficult to infer reliable information about
splashback signals from data of this kind without a complete understanding of these effects.

\section{Conclusions}\label{sec:conclusions}

In their analysis of a volume-limited sample of 8648 clusters selected by applying the redMaPPer algorithm to the SDSS/DR8 photometric data, \cite{more_detection_2016} detected strong assembly bias as a function of cluster concentration on projected scales $\SI{5}{\Mpch}< R <\SI{30}{\Mpch}$, and substantial variations in the slope of cluster projected galaxy number density profiles in the range $\SI{500}{\kpch} < R < \SI{5}{\Mpch}$ which they attributed to splashback effects. The assembly bias signal had previously been seen at lower signal-to-noise by \cite{miyatake_evidence_2016} in gravitational lensing data for the same cluster sample. By using a simplified version of the redMaPPer scheme on three orthogonal projections of publicly available galaxy catalogues from the Millennium Simulation, we  have been able to identify up to 9196 clusters of similar richness, which we classify by concentration in a similar way to the SDSS studies. This allows us to carry out analyses directly analogous to those of \cite{more_detection_2016} and \cite{miyatake_evidence_2016} and to compare with
results obtained from the full 3D information available for the simulation. This
gives considerable insight into the features seen in the SDSS analysis.

The mean projected profiles of mass and galaxy number density which we find for the simulation are very similar to those found observationally, both for the
cluster sample as a whole and for its low- and high-concentration subsamples. 
The apparent assembly bias on large scales agrees well with that observed, as does
the shape of the ratio of the low- and high-concentration profiles which rises with decreasing projected radius $R$ to a peak at the mean value of $R_c$, the limiting radius used to define clusters, before dropping precipitously to smaller scales. The
variation with $R$ of the logarithmic slope of the mean galaxy number density
profiles shows a more complex structure than in SDSS, but reproduces the main
features pointed out by  \cite{more_detection_2016}: the main minimum (the point where the profile is steepest) occurs at smaller radius than expected from the splashback studies of \cite{diemer_dependence_2014} and in addition the minima
for the low- and high-concentration subsamples rank oppositely to the splashback expectation both in depth and in radius.

The observed large-scale assembly bias is best reproduced when all red galaxies projected onto a cluster (hence within $\pm \SI{250}{\Mpch}$ in depth) are considered as potential members. The signal is slightly weaker if the maximal allowed depth offset is reduced to \SI{120}{\Mpch} and significantly weaker if it is reduced to \SI{60}{\Mpch}. Such changes have negligible effect on the logarithmic slope profiles of stacked galaxy counts. Hence projection over relatively large depths appear to be a significant factor in apparent assembly bias but not in apparent splashback features.

The above results, derived by stacking simulated clusters in projection, can be compared to results obtained from a directly analogous analysis of the full 3D data.
This shows some striking differences. The 3D assembly bias for separations  between 3 and $30\si{\Mpch}$ is considerably smaller than that seen in 2D ($b\sim 1.15$ rather than $b\sim 1.5$) and varies in the opposite way with the maximum depth offset allowed for cluster members. The peak in the ratio of the galaxy number density profiles for low- and high-concentration clusters occurs at a substantially larger radius in 3D than in 2D  ($r\sim 2.5\si{\Mpch}$ rather than $R\sim 800\si{\kpch}$). The logarithmic derivatives of the 3D mass and galaxy overdensity profiles vary more smoothly than in 2D, and show a single minimum which is at larger radius than in 2D and at the same position for the low- and high-concentration clusters. The ranking of the minima in depth remains opposite to that expected from splashback theory. (See the Appendix for a discussion of how cluster selection, scaling and stacking procedures can affect apparent splashback features). 

The effects of projection and cluster definition on stacked cluster profiles can be clarified by examining them in the two-dimensional space of projected separation and line-of-sight depth. This allows identification of the depth ranges which give rise to the difference in projected counts around low- and high-concentration clusters. As expected, the galaxy excess at small projected radius which produces the high central surface density of high-concentration clusters is made up of objects which are close to the cluster centre also in 3D. In redshift space, these excess counts appear at offsets $\sim \SI{800}{\km\per\s}$, in the wings of the cluster velocity dispersion. At projected radii $500\si{\kpch}$ $<R<2\si{\Mpch}$, much of the projected count excess around low-concentration clusters comes from galaxies offset in depth by $\sim 1\si{\Mpch}$, apparently indicating that low-concentration clusters live in richer environments than their high-concentration analogues. At larger projected separation, the galaxies responsible for the strong assembly bias signal are distributed almost uniformly over the full depth accessible to potential cluster members, showing that they are correlated with background groups preferentially projected onto the low-concentration clusters, rather than with the clusters themselves. The overall effect of projection on 2D assembly bias clearly depends strongly both on the details of cluster and concentration definition and on the accuracy of the available photometric redshifts.

At projected radii $500\si{\kpch}<R <3\si{\Mpch}$ where splashback effects are expected to be present, distant foreground and background galaxies contribute negligibly to projected cluster profiles. These are, however, strongly affected by the specific algorithms used to identify clusters and to classify them according to concentration. We demonstrate this explicitly by changing the limiting radius $R_c$ within which red galaxies are counted as cluster members. Even though we take care to adjust parameters so that the abundance and typical mass of clusters are matched for different choices of limiting radius, we find that this radius is strongly imprinted on the mean projected profiles of the resulting samples. The effects are dramatic, both on the ratio of the profiles for low- and high-concentration clusters and on the shape of the logarithmic derivative profiles for the individual subsamples. It will be difficult to obtain reliable information about splashback without detailed understanding of such effects for the particular algorithms used to select an observed cluster sample. 

\section*{Acknowledgements}

The authors thank Surhud More for useful discussions of this work.


\bibliography{phd}
\bibliographystyle{mnras}

\appendix

\section{The effect of stacking procedures on apparent splashback signal}\label{sec:mean_median_deriv}

In \autoref{sec:3d_gal_profiles} we noted that logarithmic derivative
curves for the stacked 3D mass profiles of our clusters (\autoref{fig:mat_dens3d_deriv_comp}) differ in shape, particularly in the depth of the minimum, from those shown for objects of similar mass by \cite{diemer_dependence_2014} \citepalias{diemer_dependence_2014}. A general difference in behaviour between mean and median of profile stacks was already mentioned in \citetalias{diemer_dependence_2014}. Here we investigate how the shapes of such profiles depend on the definition of the sample to be stacked and on the scaling and stacking procedures adopted.

\begin{figure}
    \centering
        \includegraphics[scale=1]{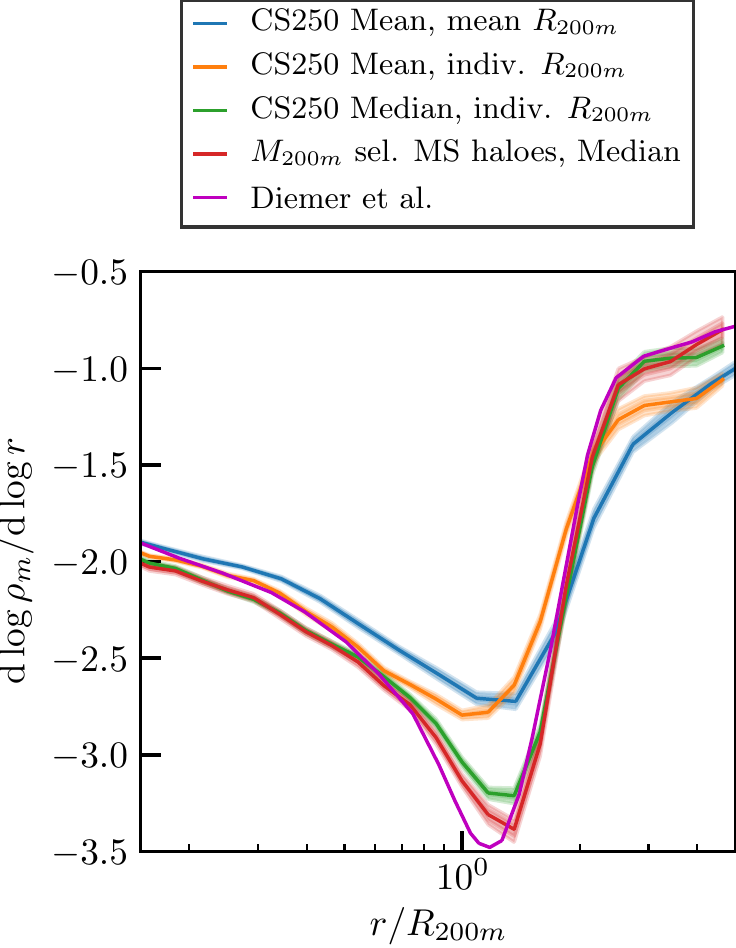}
        \caption{Logarithmic derivative curves for different definitions of the radially rescaled 3D mass density profile of simulated clusters are compared to the $z=0.25$, $2<\nu<2.5$ curve given in \protect\citetalias{diemer_dependence_2014}. For a description of the other curves the reader is referred to the text in Appendix~\protect\ref{sec:mean_median_deriv}.}\label{fig:mean_median_deriv}
    \end{figure}

In \autoref{fig:mean_median_deriv}, the purple curve is taken directly from \citetalias{diemer_dependence_2014} where it is the one labelled $z=0.25$ in the upper central panel of their Figure 4. It corresponds to haloes in a relatively narrow range of $M_{200m}$, selected at a redshift and with a mean mass which are close to those of the cluster sample analysed in this paper. \citetalias{diemer_dependence_2014} scaled the 3D mass profile of each cluster to its individual $R_{200m}$ and then constructed the stack by taking the {\it median} value of density at each $r/R_{200m}$. The logarithmic derivative of the resulting profile is the quantity
plotted. Note that it differs from the quantity plotted in \autoref{fig:mat_dens3d_deriv_comp} in that \citetalias{diemer_dependence_2014} did not
subtract the mean background density from their stack. This has a significant effect 
beyond a few Mpc.

The light blue curve in \autoref{fig:mean_median_deriv} corresponds
to our full sample CS250, stacked in the same way as in \autoref{sec:3d_gal_profiles}, i.e. we constructed a spherically averaged mass profile around the central galaxy of each cluster, we averaged these profiles directly to obtain the stack, we scaled the result by the $\langle R_{200m}\rangle$ of the stack, and we then plotted its derivative. The curve effectively corresponds to an average of the two curves shown in \autoref{fig:mat_dens3d_deriv_comp}, except for differences at large $r/R_{200m}$
due to the inclusion of the cosmic mean density. Its minimum value is about -2.7, just above the average of the values for the two curves in \autoref{fig:mat_dens3d_deriv_comp} and considerably above the value found by \citetalias{diemer_dependence_2014}.

The orange curve in \autoref{fig:mean_median_deriv} shows what happens if we scale the profile of each cluster in radius by its individual value of $R_{200m}$ 
before stacking.  This changes the shape of the curve, lowering its minimum slightly and moving it to slightly smaller radii. Not surprisingly, scaling before stacking results in a sharper transition between the one-halo and two-halo parts of the stacked profile.

If we stack these same scaled profiles by constructing their median at each $r/R_{200m}$, rather than their mean, we obtain the green curve. The minimum is
now significantly deeper, although still not as deep as that found by \citetalias{diemer_dependence_2014}. The shape of the curve outside the minimum
agrees very well with their results.

Finally, if we select halos directly from the Millennium Simulation with a narrow range of $M_{200m}$ at $z=0.24$, and we make a median stack after scaling each profile to its individual $R_{200m}$ value, then we should be reproducing the
halo selection and stacking procedures of \citetalias{diemer_dependence_2014} almost exactly. The result is shown as a red curve in \autoref{fig:mean_median_deriv}.
It now differs only slightly from the purple curve.

We suspect that these small residual discrepancies reflect differences in the effective smoothing associated with halo profile construction and differentiation. Overall, the results described here indicate that curves of this type are sensitive to
how the halos are scaled and whether a mean or median stack is constructed. The minimum logarithmic slope is particularly sensitive to these factors, and changes in shape
can also shift the position of the minimum by 10 or 20 per cent. We note that for individual observed clusters the value of $R_{200m}$ is unknown, the full 3D information is not available, and the selection and definition effects on 2D
profiles which we discuss in the main body of our paper are large compared to the effects described here.

\end{document}